\documentclass[11pt, a4paper]{article}
\pdfoutput=1

\usepackage{jcappub}
\usepackage{comment}
\usepackage{amsmath}
\usepackage{amsfonts}
\usepackage{amssymb}
\usepackage{graphicx}
\usepackage{mathrsfs}
\usepackage{latexsym}
\usepackage{color}
\usepackage{slashed, cancel}
\usepackage[normalem]{ulem} %For striking out text
\usepackage{hyperref}
\usepackage{cleveref}
\usepackage{float}
\usepackage{tikz}
\usepackage{bbold}%For identity-matrix symbol
\usepackage{soul} %For strikethrough, \st{text}
\usepackage{mathtools}% http://ctan.org/pkg/mathtools
\usepackage{support-caption} 
\usepackage{subcaption}
\usepackage{enumitem}
\usepackage{journals}
\allowdisplaybreaks

\usepackage[utf8]{inputenc} 

\hypersetup{urlcolor=blue, colorlinks=true} % Colors hyperlinks in blue - change to black if annoying

\usepackage{fancyhdr}
\numberwithin{equation}{section}

\definecolor{rossos}{rgb}{0.8,0.2,0.3}
\definecolor{bluscuro}{rgb}{0.15, 0.2, .85}
\definecolor{bluchiaro}{cmyk}{1,.3,0.,0.1}
\hypersetup{colorlinks, citecolor=bluscuro, linkcolor=bluscuro, urlcolor=bluscuro}

\makeatother   % Cancel the effect of \makeatletter
\newcommand{\GeV}{{\rm \,GeV}}
\newcommand{\TeV}{{\rm \,TeV}}

\newcommand{\mathsc}[1]{\text{\textsc{#1}}}

\newcommand{\met}{\slashed{E}_T}

\def\be   {\begin{equation}}   \def\ee   {\end{equation}}
\def\ba   {\begin{array}}      \def\ea   {\end{array}}
\def\bea  {\begin{eqnarray}}   \def\eea  {\end{eqnarray}}
\def\bal  {\begin{align}}      \def\eal  {\end{align}}
\def\bean {\begin{eqnarray*}}  \def\eean {\end{eqnarray*}}
\def\nn{\nonumber}

\begin{document}

%\today

\title{\begin{huge}
Comparing 2HDM $+$ Scalar and Pseudoscalar Simplified Models at LHC
\end{huge}}

\author[a,b]{Giorgio Arcadi,}
\author[c]{Giorgio Busoni,}
\author[c]{Thomas Hugle}
\author[c]{and Valentin Titus Tenorth}
\affiliation[a]{Dipartimento di Matematica e Fisica, Universit\'a di Roma Tre, Via della Vasca Navale 84, 00146, Roma, Italy}
\affiliation[b]{INFN Sezione Roma Tre, Italy}
\affiliation[c]{Max-Planck-Institut f\"ur Kernphysik, Saupfercheckweg 1, 69117 Heidelberg, Germany}

\emailAdd{giorgio.arcadi@uniroma3.it}
\emailAdd{giorgio.busoni@mpi-hd.mpg.de}
\emailAdd{thugle@mpi-hd.mpg.de}
\emailAdd{tenorth@mpi-hd.mpg.de}

\abstract{
In this work we compare the current experimental LHC limits of the 2HDM $+$ scalar and pseudoscalar for the $t \bar{t}$, mono-$Z$ and mono-$h$ signatures and forecast the reach of future LHC upgrades for the mono-$Z$ channel. Furthermore, we comment on the possibility, in case of a signal detection, to discriminate between the two models. The 2HDM+S and 2HDM+PS are two notable examples of the so-called next generation of Dark Matter Simplified Models. They allow for a renormalizable coupling of fermionic, Standard Model singlet, Dark Matter with a two Higgs doublet sector, through the mixing of the latter with a scalar or pseudoscalar singlet.
}

\maketitle

\section{Introduction}

The search of Dark Matter (DM) has become since some years one of the primary objectives of the LHC collaboration and represents, as well, one of the main motivations for the proposal of new collider facilities. From a theoretical perspective, it is crucial to provide an efficient interface for the interpretation of the outcome of collider searches and to enforce the complementarity with the other DM search strategies, namely Direct and Indirect Detection.

The so-called simplified models (see e.g.~\cite{Abdallah:2014hon,Arcadi:2017kky} for reviews) were born to satisfy this need. Their validity might be, however, questionable. The reason for this is twofold: on the one hand, their extreme simplicity might render these models overconstrained and on the other hand, associate them a very limited set of experimental signatures. 

For example, given the fact that in these models the DM and the mediator are singlets with respect to the SM gauge group, the typical DM related collider signature is its pair production in association with initial state radiation (ISR). From the several considered candidates for ISR, like gluons, photons and $W$, $Z$ and Higgs bosons, the strongest constraints arise from gluon initial state induced mono-jet events. This holds particularly true for the (pseudo)scalar mediator under study in this work, since gluon fusion is typically the dominant production mechanism for them at the LHC and jets face the strongest SM gauge coupling. However, it is not expected that constraints from mono-jet searches significantly improve in the near future as the are already limited by systematic uncertainties and therefore do not benefit from only increasing the total accumulated luminosity \cite{Bauer:2017ota, Aaboud:2017phn, Sirunyan:2017jix}. Thus it is worth to investigate the potential constraints from mono-$Z/h$ which, on the contrary, are expected to improve significantly in the future LHC runs.
On the theoretical side it is then necessary to elaborate appropriate models to interpret this kind of signals. In particular, the $Z$ or Higgs boson should not originate from ISR but rather be the products of a decaying mediator. This can naturally appear in models in which the mediator is not a complete SM singlet, which it also does not have to be from a theoretical perspective. Especially if the (pseudo)scalar mediator is supposed to couple to SM fermions in a gauge invariant way, it needs to be at least charged under the EW gauge group or coupled to the Higgs sector. Furthermore, as pointed out in e.g.~\cite{No:2015xqa, Bauer:2017ota, Bell:2017rgi, Abe:2018bpo,vonBuddenbrock:2016rmr,vonBuddenbrock:2018xar,Baum:2017gbj,Baum:2018zhf,Baum:2019pqc}, simplified models that lack theoretical completions might potentially lead to unreliable predictions.

For this reason the community is actively looking for a new generation of DM models~\cite{Abe:2018bpo} which can account for a rich collider phenomenology and solid theoretical predictions still within a not too large set of free parameters. 

The so called 2HDM+S and 2HDM+PS are two notable examples of this last category of models. First of all, they allow for a renormalizable coupling between a fermionic gauge singlet DM candidate and the Standard Model (SM). The DM couples, on first instance, with a (pseudo)scalar singlet. A portal with the SM sector is created by the mass mixing of the latter with the CP-even (CP-odd) states of a two doublet Higgs sector. The 2HDM+S and 2HDM+PS provide, moreover, a broad variety of collider signals, beyond simple missing energy signatures (see e.g.~\citep{No:2015xqa, Bauer:2017ota, Bell:2017rgi, Abe:2018bpo,vonBuddenbrock:2016rmr,vonBuddenbrock:2018xar,Baum:2017gbj,Baum:2018zhf,Baum:2019pqc}).

The aim of this paper is to study some of these particular signatures, namely $t\bar t$ resonances, mono-$Z$ and mono-$h$. Although, all of these signatures appear in both models, there can be a sizeable difference for the expected signal rates in the pseudoscalar and scalar model. Therefore, in our work we characterize the similarities and differences of the different signatures in the two models, by looking at the limits derived from current LHC data and analyses. In addition, we also look at the reach of the mono-$Z$ channel for future luminosities and comment on the possibility to discriminate between the 2HDM+S and 2HDM+PS models in case of a future signal detection.

The paper is structured as follows. In the next section we will illustrate, from a theoretical perspective, the two models. Section~3 is, instead, devoted to a brief overview of their present phenomenological constraints. The results of our numerical study will be presented in section~4. In Section~5 we finally draw our conclusions.

\section{2HDM plus Scalar/Pseudoscalar: Model Description}
\label{sec:modeldescription}

The models we analyse are 2HDMs containing an additional singlet scalar particle. These models have been widely discussed in litterature, e.g.~\citep{Ipek:2014gua, Bell:2016ekl, Bell:2017rgi, Sanderson:2018lmj, Arcadi:2017wqi, Arcadi:2018pfo, Bauer:2017ota}.

\subsection{Scalar Potential}
\label{sec:potential}

The most general scalar potential we consider is
\begin{equation}
    V(\Phi_1,\Phi_2,S) = V_{\mathsc{2hdm}}(\Phi_1,\Phi_2) + V_S(S) + V_{S\mathsc{2hdm}}(\Phi_1,\Phi_2,S), \label{eq:fullpotential}
\end{equation}
where $V_{\mathsc{2hdm}}$ is the standard 2HDM potential, $V_S$ is the potential of the scalar singlet and $V_{S\mathsc{2hdm}}$ is the potential involving interaction terms between the scalar singlet and the doublets. We start by reviewing the 2HDM part of the potential. In 2HDMs, one has two doublets with identical charges, therefore having the freedom to choose a specific base $\Phi_1,\Phi_2$ in terms of which to write the potential. A generic $SU(2)$ change of base,
\begin{align}
    \left(\begin{array}{c}
        \Phi_1 \\
        \Phi_2 \\
    \end{array} \right)\rightarrow \left(
    \begin{array}{c}
        \Phi_1^{'} \\
        \Phi_2^{'} \\
    \end{array} \right)= U \left(
    \begin{array}{c}
        \Phi_1 \\
        \Phi_2 \\
    \end{array}\right)
\end{align}
where $U$ is an $SU(2)$ matrix, will result in a potential giving the same physics but with different coefficients for the different terms. While it would be possible to get rid of residual freedom through basis independent methods, see e.g.~\cite{Davidson:2005cw,Trautner:2018ipq,Ivanov:2019kyh}, we will nevertheless consider, in this paper, two reference bases in which to write the potential. The first one is what we will call the flavour base. Assuming the Yukawa interactions of the 2HDM are either of type I, II, X, Y or Inert, the basis $\Phi_1,\Phi_2$ is defined in terms of which of the doublets interacts with the various fermions, and is therefore well defined. This basis is very useful to analyse the interactions of the scalars with fermions. The second basis we choose is the so-called Higgs basis, where one of the two doublets has no vacuum expectation value (vev). When writing the potential, we will label $\lambda_i, \, M_{ij}$ the coefficients of the potential in the flavour base, and we will use $\hat{\lambda}_i,\hat{M}_{ij}$ to label the coefficients of the potential in the Higgs basis. The doublets in the flavour basis will be labelled $\Phi_1,\Phi_2$, while the doublets in the Higgs basis will be labelled $\Phi_h,\Phi_H$. The 2HDM potential reads in the flavour basis
\begin{align}
\label{eq:2HDM_potential}
    V_{\mathsc{2hdm}}(\Phi_1,\Phi_2) &= M_{11}^2 \Phi_1^\dagger \Phi_1 + M_{22}^2 \Phi_2^\dagger \Phi_2 +  (M_{12}^2 \Phi_2^\dagger \Phi_1 + h.c.) + \frac{\lambda_1}{2} (\Phi_1^\dagger \Phi_1)^2 + \frac{\lambda_2}{2} (\Phi_2^\dagger \Phi_2)^2 \nonumber \\
    &\quad + \lambda_3 (\Phi_1^\dagger \Phi_1)(\Phi_2^\dagger \Phi_2) + \lambda_4 (\Phi_2^\dagger \Phi_1)(\Phi_1^\dagger \Phi_2) + \frac{1}{2}\left(\lambda_5 (\Phi_2^\dagger \Phi_1)^2 + h.c.\right),
\end{align}
where a $\mathcal{Z}_2$ symmetry $\Phi_1 \rightarrow \Phi_1,\ \Phi_2 \rightarrow -\Phi_2$ has been imposed to suppress flavour changing neutral currents, thus removing the additional terms $\lambda_{6,7}$ containing three $\Phi_1$ doublets and one $\Phi_2$ doublet and vice versa. We still allow for the soft-breaking term $M_{12}^2$ that is necessary to have a scalar mass spectrum with the desired physical features. We also assume that the potential does not break CP explicitly. This means  $(M_{12}^2)^2$ and $\lambda_5$ need to have the same phase (up to a negative relative sign), which can be reabsorbed into $\Phi_2$ by a field redefinition. Therefore we will assume that the parameters $\lambda_i,M_{ij}$ are all real. In this case, the $SU(2)$ transformation freedom for the doublets reduces to $SO(2)$ rotations:
\begin{align}
    \left(\begin{array}{c}
        \Phi_1\\
        \Phi_2 \\
    \end{array}\right)\to \left(
    \begin{array}{c}
        \Phi_1^{'}\\
        \Phi_2^{'}\\
    \end{array}\right)= \left(
    \begin{array}{cc}
        \cos\beta & \sin\beta\\
        -\sin\beta & \cos\beta\\
    \end{array}\right) \left(
    \begin{array}{c}
        \Phi_1 \\
        \Phi_2 \\
    \end{array}\right)
\end{align}
The most general expressions for the other two terms in Eq.~\eqref{eq:fullpotential} are
\begin{align}
    V_S(S) &= \frac{1}{2} M_{SS}^2 S^2 + \frac{1}{3} \mu_S S^3 + \frac{1}{4} \lambda_S S^4, \\
    V_{S\mathsc{2hdm}}(\Phi_1,\Phi_2,S) &= \mu_{11S}(\Phi_1^\dagger \Phi_1)S + \mu_{22S}(\Phi_2^\dagger \Phi_2)S + (\mu_{12S} \Phi_2^\dagger \Phi_1 S + h.c.) \nonumber\\
    &\quad + \frac{\lambda_{11S}}{2}(\Phi_1^\dagger \Phi_1)S^2 +  \frac{\lambda_{22S}}{2}(\Phi_2^\dagger \Phi_2)S^2 + \frac{1}{2}(\lambda_{12S} \Phi_2^\dagger \Phi_1 S^2 + h.c.).
\label{eq:vs2hdm}
\end{align}
The $\mathcal{Z}_2$ symmetry might be also extended to these terms of the potential by defining suitable transformation properties for $S$. By choosing $S\rightarrow -S$ the terms $\mu_S,\, \mu_{11S},\, \mu_{22S}$, would be forbidden, however they would still be allowed in case of soft-breaking of the $\mathcal{Z}_2$. The $\lambda_{12S}$ term on the other hand is forbidden due to the assumed $\mathcal{Z}_2$ symmetry of the Higgs fields. All these terms would have, however, a negligible impact for the phenomenology discussed in this work, since they would affect only the scalar trilinear and quartic interactions. Consequently, in line with \citep{Bauer:2017ota}, we will not include them in our analysis.

In Eq.~\eqref{eq:vs2hdm}, all coefficients need to be real, except for $\lambda_{12S}$ and $\mu_{12S}$. As we have set $\lambda_{12S} = 0$, the only remaining physical phase is the one of $\mu_{12S}$. To conserve CP, this coefficient must be either purely real or imaginary (after rotating the 2HDM doublets to have all phases contained in the 2HDM potential reabsorbed). In the first case, mixing between the CP-even scalar states contained in the $\Phi_1$ and $\Phi_2$ doubles and the singlet scalar would be obtained. In the latter case instead, the singlet state would mix with the neutral CP-odd states contained in the doublets. 
 
The 2HDM+PS, defined in \citep{Ipek:2014gua}, and discussed in \citep{Bauer:2017ota}, is obtained using the latter option. Relabeling $S$ to $P$ to evidence more explicitly its CP-odd nature, the resulting potential is
\begin{equation}
    V(\Phi_1,\Phi_2,P) = V_{\mathsc{2hdm}}(\Phi_1,\Phi_2) + V_P(P) + V_{P\mathsc{2hdm}}(\Phi_1,\Phi_2,P), 
\label{eq:potentialp}
\end{equation}
where $V_{\mathsc{2HDM}}$ is the one given in Eq.~\eqref{eq:2HDM_potential} while
\begin{align}
    V_P(P) &= \frac{1}{2}\, M_{PP}^2 P^2 + \frac{1}{4}\, \lambda_P P^4, \\
    V_{P\mathsc{2hdm}}(\Phi_1,\Phi_2,P) &= \frac{\lambda_{11P}}{2}\, (\Phi_1^\dagger \Phi_1)P^2 +  \frac{\lambda_{22P}}{2}\, (\Phi_2^\dagger \Phi_2)P^2 + \mu_{12P} P(i \Phi_1^\dagger \Phi_2 + h.c.),
\label{eq:p2hdm}
\end{align}
with
\begin{align}
    \Phi_i = \left(
     \begin{array}{cc}
      \Phi_i^+ \\
      \frac{v_i + \rho_i + i \eta_i}{\sqrt{2}}\\
     \end{array}
    \right), \quad P = \eta_3,
\end{align}
and $v_{1,2}$ are usually parametrized in terms of
\begin{equation}
    \tan\beta =  \frac{v_2}{v_1}, \;\;\;\; {\textrm{with}} \;\;\;\; v_1^2+v_2^2 = v^2.
\end{equation}
$\rho_{1,2}$ are CP-even scalar particles, and $\eta_{1,2,3}$ are CP-odd scalar particles, but not necessarily mass eigenstates.

It is convenient to rotate $\{\Phi_1, \Phi_2\}$ to the Higgs basis $\{\Phi_h, \Phi_H \}$, where $\langle \Phi_H \rangle = 0$ and $\langle \Phi_h \rangle = v \approx 246$ GeV. The potential then reads
\begin{equation}
    \hat{V}(\Phi_h,\Phi_H,P) = \hat{V}_{\mathsc{2hdm}}(\Phi_h,\Phi_H) + \hat{V}_P(P) + \hat{V}_{P\mathsc{2hdm}}(\Phi_h,\Phi_H,P),
\label{eq:potentialph}
\end{equation}
where
\begin{align}
    \hat{V}_{\mathsc{2hdm}}(\Phi_h,\Phi_H) &= \hat{M}_{hh}^2 \Phi_h^\dagger \Phi_h + \hat{M}_{HH}^2 \Phi_H^\dagger \Phi_H + (\hat{M}_{hH}^2 \Phi_H^\dagger \Phi_h + h.c.) + \frac{\hat{\lambda}_h}{2} (\Phi_h^\dagger \Phi_h)^2 + \frac{\hat{\lambda}_H}{2} (\Phi_H^\dagger \Phi_H)^2 \nonumber\\
    &\quad + \hat{\lambda}_3 (\Phi_h^\dagger \Phi_h)(\Phi_H^\dagger \Phi_H) + \hat{\lambda}_4 (\Phi_H^\dagger \Phi_h)(\Phi_h^\dagger \Phi_H) + \frac{\hat{\lambda}_5}{2} \left( (\Phi_H^\dagger \Phi_h)^2 + h.c.\right)\nonumber\\
    &\quad + \hat{\lambda}_6 \Phi_h^\dagger\Phi_h\left(\Phi_H^\dagger \Phi_h + h.c.\right)+\hat{\lambda}_7 \Phi_H^\dagger\Phi_H\left(\Phi_H^\dagger \Phi_h + h.c.\right),\\
    \hat{V}_P(P) &= \frac{1}{2}\, \hat{M}_{PP}^2 P^2 + \frac{\hat{\lambda}_P}{4}\, P^4,\\
    \hat{V}_{P\mathsc{2hdm}}(\Phi_h,\Phi_H,P) &= \frac{\hat{\lambda}_{HHP}}{2}\, \Phi_H^\dagger \Phi_H P^2 + \frac{\hat{\lambda}_{hhP}}{2}\ \Phi_h^\dagger \Phi_h P^2 + \frac{\hat{\lambda}_{hHP}}{2} P^2 (\Phi_H^\dagger \Phi_h  + h.c.)\nonumber\\
    &\quad + \mu_{12P}P\left(i\Phi_h^\dagger\Phi_H + h.c.\right). 
\end{align}
The two Higgs doublets are then defined by
\begin{align}
    \Phi_h &= \cos\beta \ \Phi_1 + \sin\beta \ \Phi_2 = \left(
     \begin{array}{cc}
      G^+ \\ \frac{v + \hat{\rho}_1 + i G^0}{\sqrt{2}} \\
     \end{array}
    \right),\label{eq:palignh}\\
    \Phi_H &= -\sin\beta \ \Phi_1 + \cos\beta \ \Phi_2 = \left(
     \begin{array}{cc}
      H^+ \\ \frac{ \hat{\rho}_2 + i \hat{\rho}_3}{\sqrt{2}}\\
     \end{array} \right),
\end{align}
where $G^\pm,\, G^0$ are the SM Goldstone bosons, and $H^\pm$ is the 2HDM charged scalar. All these particles are mass eigenstates. The particles $\hat{\rho}_{1,2}$ are CP-even scalars that are linear combinations of the standard model Higgs boson $h$ and an additional heavy scalar $H$. $\hat{\rho}_{3},\eta_3$ are, instead, CP-odd and their combination will originate mass eigenstates which will be labelled $a$ and $A$ with, in general, $M_a < M_A$.
Note that, in general, the terms $\lambda_6,\lambda_7,\lambda_{hHP}$ can arise when changing base. Moreover, the coefficient of the term $P\left(i\Phi_h^\dagger\Phi_H + h.c.\right)$ does not change when changing base.

For the scalar case, it is possible to proceed in the same way, assuming $\mu_{12S}$ to be purely real. However, in the case of the scalar, there is also another way to obtain a mixing between the singlet and the CP-even scalars of the doublets. One can in fact assume that the singlet develops a vev. This is the approach taken by \citep{Bell:2016ekl,Bell:2017rgi}.
In this approach, one assumes that the scalar potential has a spontaneously broken $\mathcal{Z}_2$ symmetry w.r.t.~which only the particle $S$ is odd, that gets rid of all cubic terms, independently of the basis in which the potential is written in. This may arise naturally, for example, in the case where $S$ is part of a complex scalar charged under a dark $U(1)$ gauge group. In~\citep{Bell:2017rgi}, the authors write the potential directly in the Higgs basis as
\begin{equation}
    \hat{V}(\Phi_h,\Phi_H,S) = \hat{V}_{\mathsc{2hdm}}(\Phi_h,\Phi_H) + \hat{V}_S(S) + \hat{V}_{S\mathsc{2hdm}}(\Phi_h,\Phi_H,S), 
\label{eq:potentialsh}
\end{equation}
where
\begin{align}
    \hat{V}_{\mathsc{2hdm}}(\Phi_h,\Phi_H) &= \hat{M}_{hh}^2 \Phi_h^\dagger \Phi_h + \hat{M}_{HH}^2 \Phi_H^\dagger \Phi_H + (\hat{M}_{hH}^2 \Phi_H^\dagger \Phi_h + h.c.) + \frac{\hat{\lambda}_h}{2} (\Phi_h^\dagger \Phi_h)^2 + \frac{\hat{\lambda}_H}{2} (\Phi_H^\dagger \Phi_H)^2 \nonumber\\
    &\quad + \hat{\lambda}_3 (\Phi_h^\dagger \Phi_h)(\Phi_H^\dagger \Phi_H) + \hat{\lambda}_4 (\Phi_H^\dagger \Phi_h)(\Phi_h^\dagger \Phi_H) + \frac{\hat{\lambda}_5}{2} \left( (\Phi_H^\dagger \Phi_h)^2 + h.c.\right),\\
    \hat{V}_S(S) &= \frac{1}{2} \hat{M}_{SS}^2 S^2 + \frac{\hat{\lambda}_S}{4} S^4, \\
    \hat{V}_{S\mathsc{2hdm}}(\Phi_h,\Phi_H,S) &= \frac{\hat{\lambda}_{HHS}}{2}\, \Phi_H^\dagger \Phi_H S^2 + \frac{\hat{\lambda}_{hhS}}{2}\, \Phi_h^\dagger \Phi_h S^2 + \frac{1}{2} \left(\hat{\lambda}_{hHS} \Phi_H^\dagger \Phi_h S^2 + h.c.\right),
\end{align}
with
\begin{align}
    \Phi_h &= \left(
    \begin{array}{cc}
     G^+ \\ \frac{v + \hat{\rho}_1 + i G^0}{\sqrt{2}} \\
    \end{array} \right),\\
    \Phi_H &=  \left(
    \begin{array}{cc}
     H^+ \\\frac{ \hat{\rho}_2 + i A}{\sqrt{2}} \\
    \end{array} \right),\\
    S &= v_s + \hat{\rho}_3 .
\label{eq:salignh}
\end{align}
This time, on top of the SM Goldstone bosons $G$ and the charged scalar $H^\pm$, there is a single CP-odd scalar, $A$, and three CP-even scalars, $\hat{\rho}_{1,2,3}$. Those mix together, and are linear combinations of $h$ and $S_{1,2}$, which are two new CP-even scalar particles.     
Note that, in this potential written in the Higgs base the terms $\hat{\lambda}_{6,7}$ are absent, contrary to the pseudoscalar case, where changing to the Higgs base switches these terms on in general.

\subsection{Mass Spectrum and Alignment Limit}
\label{sec:alignment}

The $M_h=125\GeV$ Higgs boson is experimentally observed to be very similar to the SM one. To avoid most of the constraints from Higgs physics, we choose to work in the so called alignment limit, i.e. we impose specific relations among the parameters of the scalar potential such that the mixing between the $\hat{\rho}_1$ and $\hat{\rho}_2$ is negligible and $\hat{\rho}_1$ is always identified as the experimentally observed 125\,GeV SM-like CP-even Higgs. 

In the pseudoscalar model, the alignment limit can be achieved by assuming a specific value for $\beta$, namely
\begin{equation}\label{eq:aligntb}
    \cos 2\beta = -\frac{\lambda_1-\lambda_2}{\lambda_1+\lambda_2-2(\lambda_3+\lambda_4+\lambda_5)}
\end{equation}
Alternatively, one can assume the presence of the CP2 symmetry \citep{Dev:2014yca}, that gives the following relations between the parameters
\begin{equation}\label{eq:alignCP2}
    \lambda_1 = \lambda_2 = \lambda_3+\lambda_4+\lambda_5 .
\end{equation}
In the latter approach, one has the advantage that, when switching base, one has $\lambda_1 = \hat{\lambda}_h$, $\lambda_2 = \hat{\lambda}_H$, $\lambda_i = \hat{\lambda}_i$ for $i = 3, 4, 5$ and $\hat{\lambda}_{6,7} = 0$.

In case of the scalar model, one also needs the additional assumption
\begin{equation} \label{eq:alignscalar}
    \hat{\lambda}_{hhS} = 0 .
\end{equation}
This additional condition arises because of the different choice on how the mixing is achieved, i.e. with non-zero vev for $S$.

Assuming either Eq.~\eqref{eq:aligntb} or \eqref{eq:alignCP2}, the mass matrix of the model is block diagonal. In the pseudoscalar model, it is made of a diagonal block containing two zero eigenvalues, corresponding to the Goldstone bosons $G^0,\, G^\pm$, and the masses $M_h^2,\, M_{H^\pm}^2,\, M_H^2$, plus a two-dimensional block. Diagonalizing the block, one gets the masses and the mixing angle of the 2 CP-odd scalars 
\begin{equation}
    \sin 2\theta = \frac{2\,v\,\mu_{12P}}{M_A^2-M_a^2} \,.
\end{equation}
In our convention, similar to other authors \citep{Bauer:2017ota,Bell:2017rgi}, we are concentrating on the case $M_a < M_A$.
The original set of parameters $\lambda_{1,2,4,5},\, M_{11}^2,\, M_{22}^2,\, M_{12}^2,\, M_{PP}^2,\, \mu_{12P}$ can therefore be expressed in terms of $M_h,\, M_H,\, M_{H^\pm},\, M_A,\, M_a,\, \theta,\, \tan\beta,\, v$ together with the alignment condition. The parameters $\lambda_3,$ $\lambda_{11P}$, $\lambda_{22P},$ $\lambda_P$ remain free if the alignment condition Eq.~\eqref{eq:aligntb} is chosen, while when choosing Eq.~\eqref{eq:alignCP2} only $\lambda_{11P},\, \lambda_{22P},\, \lambda_P$ remain free, while the value of $\lambda_3$ gets fixed. 

In the scalar model instead, the mass matrix is made of a diagonal block containing two zero eigenvalues, corresponding to the Goldstone bosons $G^0,\, G^\pm$, and the masses $M_h^2,\, M_{H^\pm}^2,\, M_A^2$, plus a two-dimensional block. Diagonalizing the block, one gets the masses and the mixing angle of the two CP-even scalars
\begin{equation}
    \sin2\theta = \frac{2 \hat{\lambda}_{hHs}\, v\, v_S }{M_{S_1}^2-M_{S_2}^2} \,.
\end{equation}
Also for the scalar case, we will concentrate on $M_{S_2} < M_{S_1}$.
Again, one can exchange the set of parameters $\hat{\lambda}_{h,H,4,5},\, M_{11}^2,\, M_{22}^2,\, M_{12}^2,\, M_{SS}^2,\, \mu_{12S}$ for the set $M_h$, $M_{S_1}$, $M_{H^\pm}$, $M_A$, $M_{S_2}$, $\theta$, $\tan\beta$, $v$, together with the alignment condition. The additional alignment condition Eq.~\eqref{eq:alignscalar} sets $\hat{\lambda}_{hhS}$. When choosing alignment condition Eq.~\eqref{eq:aligntb}, the parameters $\hat{\lambda}_3,\, \hat{\lambda}_{HHS},\, \lambda_S$ remain free, while when choosing alignment condition Eq.~\eqref{eq:alignCP2}, only $\hat{\lambda}_{HHS},\, \lambda_S$ remain free.

To avoid having the couplings $\lambda_{i}$ varying in an uncontrolled way when varying $\tan\beta$, throughout the rest of the paper we will adopt the alignment condition Eq.~\eqref{eq:alignCP2}, together with the mass degenerate assumption $M_H = M_{H^\pm} = M_A > M_a$ or $M_A = M_{H^\pm} = M_{S_1} > M_{S_2}$, respectively for the pseudoscalar/scalar model. Moreover, to get a reasonable comparison between the pseudoscalar and the scalar model\footnote{We remind the reader that the scalar model has the additional alignment constraint Eq.~\eqref{eq:alignscalar} that sets $\hat{\lambda}_{11S}$.} and to avoid variations with $\tan\beta$, we decide to set
\begin{align}
    \hat{\lambda}_{hhS} &= \hat{\lambda}_{HHS} = 0,\\
    \hat{\lambda}_{hhP} &= \hat{\lambda}_{HHP} = 0,
\end{align}
or equivalently
\begin{align}
    \lambda_{11S} &= \lambda_{22S} = 0,\\
    \lambda_{11P} &= \lambda_{22P} = 0,
\end{align}
for the rest of the paper. This is most relevant for the $hAa$ vertex ($h S_1 S_2$ vertex for the scalar model), which appears in mono-$h$ processes and depends on
\begin{align}
    \hat{\lambda}_{hhS} &= \lambda_{11S}\cos^2\beta+\lambda_{22S}\sin^2\beta,\\
    \hat{\lambda}_{hhP} &= \lambda_{11P}\cos^2\beta+\lambda_{22P}\sin^2\beta.
\end{align}
Finally, the parameters $\lambda_{P,S}$ are not relevant for our signatures.

\subsection{Yukawa Sector}
\label{sec:yukawa}

The Yukawa interactions of the SM fermions with the Higgs doublets can be expressed as
\begin{equation}
    L_{\text{Yukawa}} = - \sum_{n=1,2} \left(Y_{n,ij}^u \bar{Q}_L^i u_R^j \widetilde{\Phi}_n + Y_{n,ij}^d \bar{Q}_L^i d_R^j \Phi_n + Y_{n,ij}^l \bar{L}_L^i l_R^j \Phi_n + h.c. \right).
\label{eq:yukawah1h2}
\end{equation}
As in standard 2HDMs, we shall need to choose Yukawa structures that keep potentially dangerous flavour violating processes under control. We outline the possibilities below, and explore the DM phenomenology of these choices in Sec.~\ref{sec:direct} by determining direct detection constraints.

Rewriting Eq.~\eqref{eq:yukawah1h2} in the Higgs basis we have
\begin{equation}
    L_{\rm Yukawa} = - \sum_{n=h,H} \left(Y_{n,ij}^u \bar{Q}_L^i u_R^j \widetilde{\Phi}_n + Y_{n,ij}^d \bar{Q}_L^i d_R^j \Phi_n + Y_{n,ij}^l \bar{L}_L^i l_R^j \Phi_n + h.c. \right),
\label{eq:yukawahH}
\end{equation}
where the matrices $Y_{h,ij}^{u,d,l}$ have to be the SM Yukawa
matrices, namely
\begin{equation}
    Y_h^i \equiv Y_{\mathsc{sm}}^i,
\end{equation}
while the Yukawa matrices of the additional doublet are assumed to be proportional to the SM ones
\begin{equation}
    Y_H^i = \epsilon_i Y_{\mathsc{sm}}^i,
\end{equation}
where the $\epsilon_i$ are Yukawa scaling factors, with $i=u,d,l$. This Yukawa structure is the so-called Aligned Yukawa model \citep{Pich:2009sp, Tuzon:2010vt, Pich:2010ic, Penuelas:2017ikk, Gori:2017qwg,Rodejohann:2019izm,Chulia:2019evx}, which satisfies Natural Flavour Conservation. In special cases where the $\epsilon_i$ satisfy certain relationships, the Aligned Yukawa structure can correspond to one of the $\mathcal{Z}_2$ symmetric Yukawa structures (type I, II, X or Y), as shown in Tab.~\ref{tab:coeffs}. As we will only probe values of $\tan\beta\le3$, our constraints will be valid for all Yukawa structures included in Tab.~\ref{tab:coeffs}, except the Inert one.

The mass spectrum of the model contains an additional fermion, the DM candidate. Assuming it to be a SM singlet, it will couple, in the interactions basis, only with the scalar,
\begin{equation}
    \mathcal{L}_\mathsc{dm} = -y_\chi^S\, S\, \bar{\chi} \chi,
\end{equation}
or the pseudoscalar singlet
\begin{equation}
    \mathcal{L}_\mathsc{dm} = -y_\chi^P\, P\, \bar{\chi} \gamma^5 \chi.
\end{equation}
Due to the mass mixing between the SU(2) singlet and doublet scalars, the DM will in any case couple to the CP-even or CP-odd physical states according to whether we will consider the 2HDM+S or 2HDM+PS model. 

\paragraph{Type I, II, X and Y}
\label{sec:type12}

To suppress flavour-changing neutral currents in 2HDM, it is possible to assume the presence of a $\mathcal{Z}_2$ symmetry on the Yukawa sector, allowing only one of the two doublets $\Phi_{1,2}$ to couple to a certain type of quarks and leptons. This hypothesis goes under the name of natural flavour conservation (NFC). The presence of the charged scalar $H^\pm$ still allows FCNC to appear at loop level. Loop generated FCNC allow to set limits on $\tan\beta$ and the charged scalar mass also in the case of Higgs-alignment, that is weakly constrained by many other Higgs physics observables.
Each possible assignment of the $\mathcal{Z}_2$ charges results in a different type of 2HDM. These types are listed in Tab.~\ref{tab:coeffs}.

\begin{table}[tb]\centering
\begin{tabular}{c|c|c|c}
    Model & $\epsilon_d$ & $\epsilon_u$ & $\epsilon_l$\\ \hline
    Type I & $\cot\beta$ & $\cot\beta$ & $\cot\beta$\\
    Type II & $-\tan\beta$ & $\cot\beta$ & $-\tan\beta$\\
    Type X & $\cot\beta$ & $\cot\beta$ & $-\tan\beta$\\
    Type Y & $-\tan\beta$ & $\cot\beta$ & $\cot\beta$\\
    Inert & 0 & 0 & 0
\end{tabular}
\caption{Values of the coefficients $\epsilon_{u,d,l}$ which correspond to models with discrete ${\cal Z}_2$ symmetries.}
\label{tab:coeffs}
\end{table}

\subsection{Decay Widths and Branching Ratios}
\label{sec:decays}

In this section we compare the branching ratios (BR) of the four neutral spin-0 states in the 2HDM+PS and 2HDM+S for the later discussion. We give analytic expressions of the dominant decay widths for those and the charged scalars in App.~\ref{sec:Gamma_app}. For all plots and interpretations we used the parameter values from Eq.~\eqref{eq:parameters} and fixed $M_A=500$\,GeV and $\tan\beta=1$ such that the findings are applicable to all types of 2HDMs (besides the Inert one).

\begin{figure}    \centering
    \includegraphics[width=\textwidth]{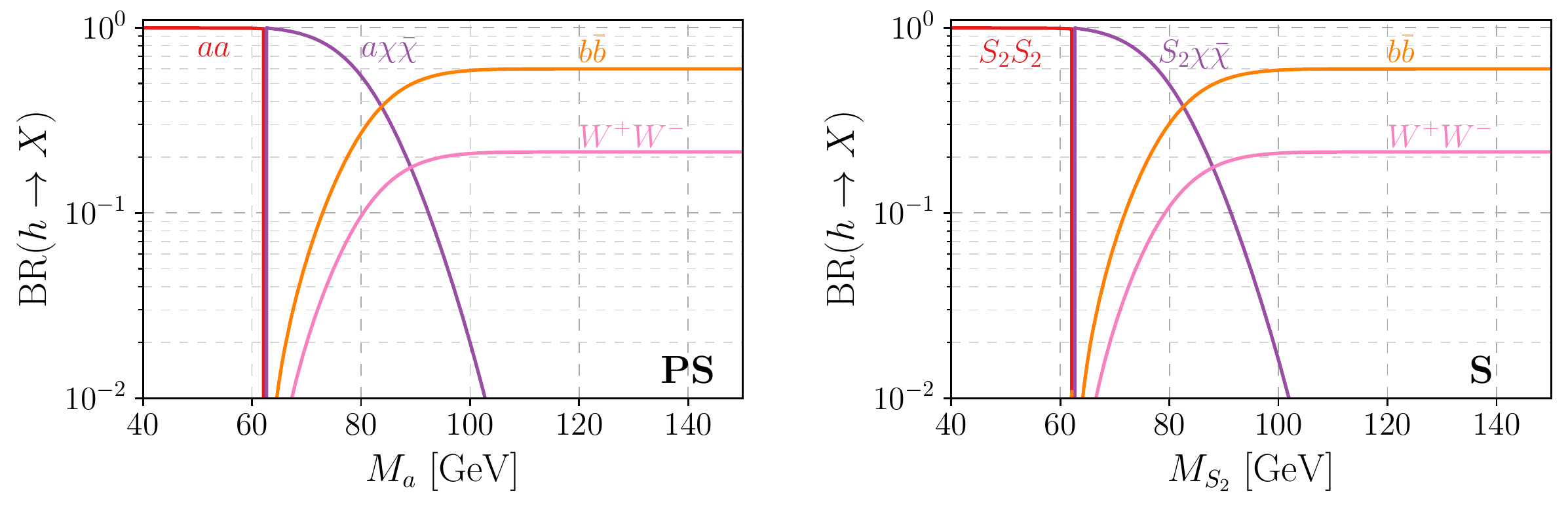}
    \caption{Dominant branching ratios of the SM Higgs-like scalar $h$ for $\tan\beta=1$, $m_\chi=10$~GeV and other parameter values as given in Eq.~\eqref{eq:parameters} in the 2HDM+PS (left) and 2HDM+S (right).}
    \label{fig:BRh}
\end{figure}

As mentioned, in the alignment limit the couplings of $h$ to the SM fields substantially coincide with the ones expected for the SM Higgs boson. However, its total width can deviate with respect to the SM prediction, because of the eventual presence of additional decay channels. The most relevant, if kinematically allowed, is the one into a pair of $a$ or $S_2$ states, respectively. Even for mediator masses above half of the SM Higgs mass, the three body decay $a\chi\bar\chi$, or $S_2\chi\bar\chi$ respectively, can be sizeable and even dominate for small enough DM masses, see Fig.~\ref{fig:BRh} (for brevity only the two dominating SM decays are shown).

\begin{figure}    \centering
    \includegraphics[width=\textwidth]{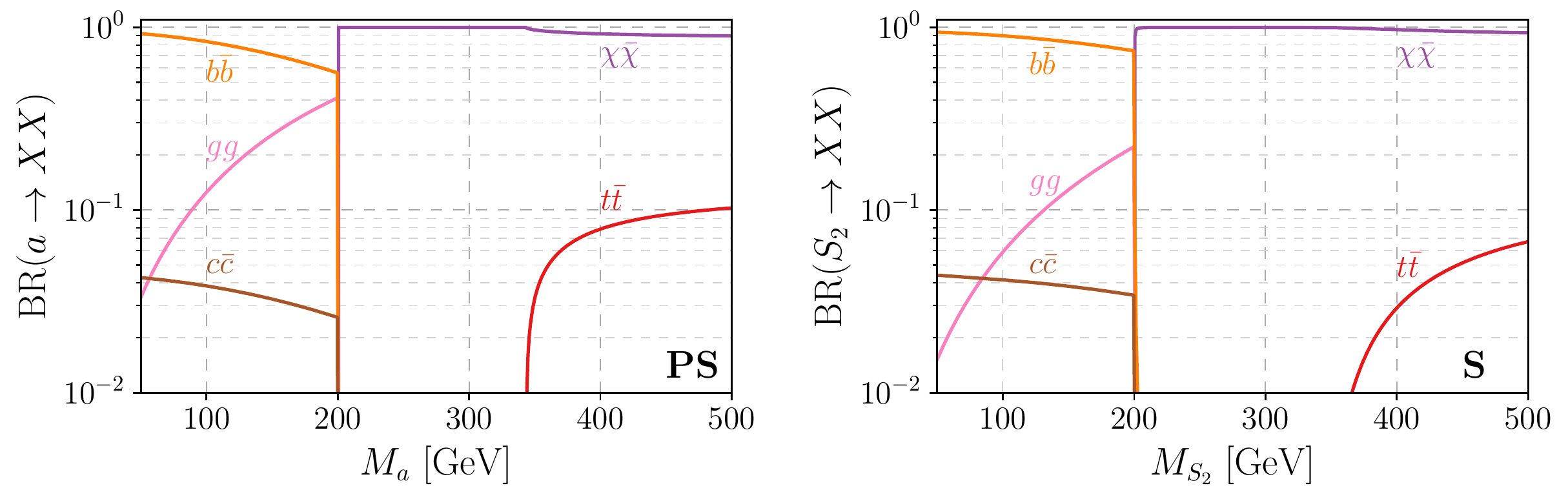}
    \caption{Dominant branching ratios of the light pseudoscalar $a$ in the 2HDM+PS (left) and the light scalar $S_2$ in the 2HDM+S (right) for $\tan\beta=1$, $m_\chi=100$~GeV (to show more decay channels), $M_{H/S_1} = M_{H^\pm} = M_A = 500$~GeV and other parameter values as given in Eq.~\eqref{eq:parameters}.}
    \label{fig:BRas}
\end{figure}

The width of the additional light (pseudo)scalar shows similar behavior in both models. It is dominated by the $\chi\bar\chi$ channel, if kinematically accessible, even if the decay to $t\bar t$ is allowed, see Fig.~\ref{fig:BRas}. The other decay channels arise from mixing with the corresponding doublet state and are therefore suppressed by $\sin^2\theta$.

\begin{figure}    \centering
    \includegraphics[width=\textwidth]{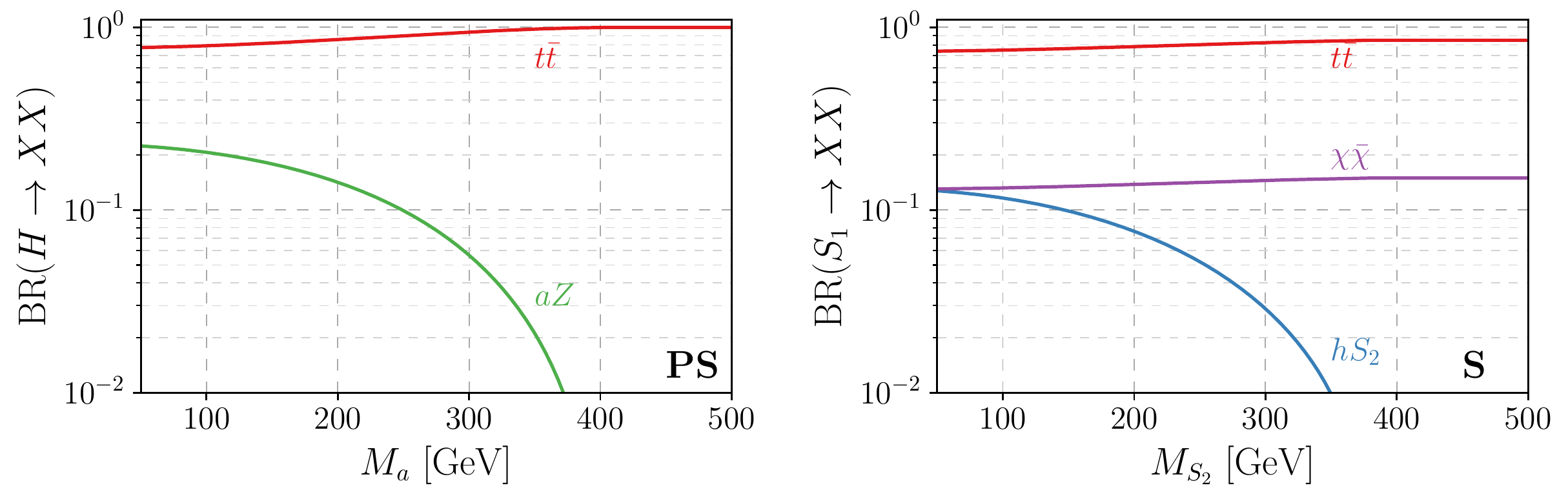}
    \caption{Dominant branching ratios of the heavy scalar $H/S_1$ for $\tan\beta=1$, $M_{H/S_1} = M_{H^\pm} = M_A = 500$~GeV, $m_\chi=10$~GeV and other parameter values as given in Eq.~\eqref{eq:parameters} in the 2HDM+PS (left) and 2HDM+S (right).}
    \label{fig:BRH}
\end{figure}

\begin{figure}    \centering
    \includegraphics[width=\textwidth]{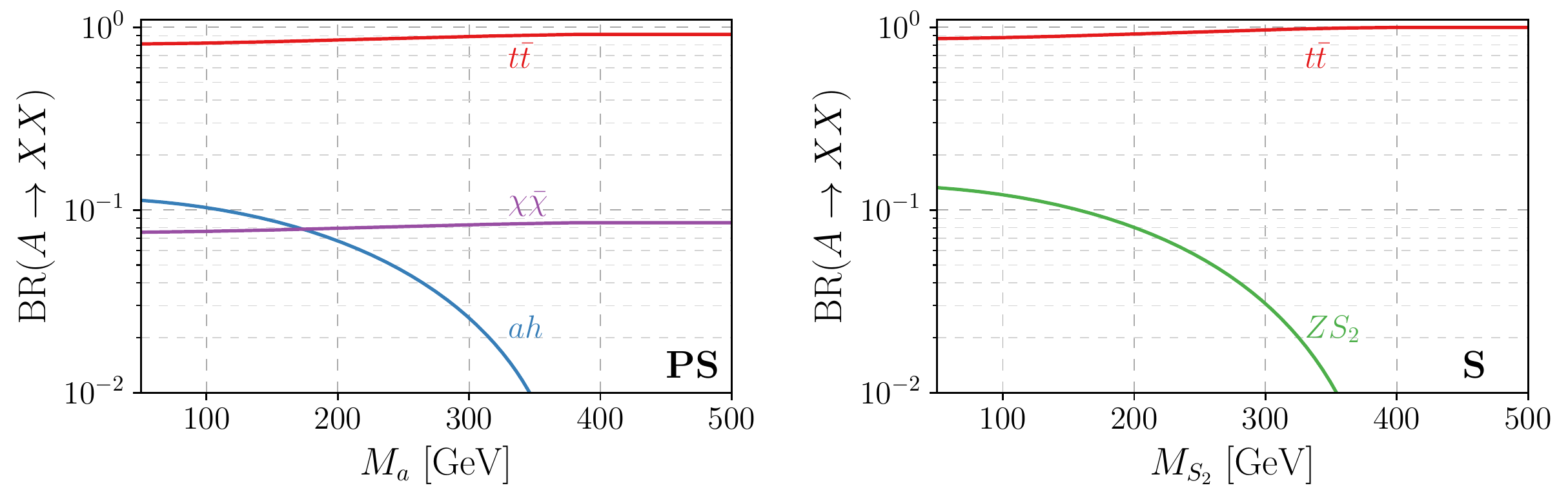}
    \caption{Dominant branching ratios of the heavy pseudoscalar $A$ for $\tan\beta=1$, $m_\chi=10$~GeV, $M_{H/S_1} = M_{H^\pm} = M_A = 500$~GeV and other parameter values as given in Eq.~\eqref{eq:parameters} in the 2HDM+PS (left) and 2HDM+S (right).}
    \label{fig:BRA}
\end{figure}

The decay channels of the heavy scalars $H/S_1$, see Fig.~\ref{fig:BRH}, and pseudoscalars $A$, see Fig.~\ref{fig:BRA}, are exchanged in the 2HDM+PS and S. All four are dominated by the decay to top quarks.
In addition the heavy pseudoscalar $A$ in the 2HDM+S and the heavy scalar $H$ in the 2HDM+PS decay to $ZS_2$, or $Za$ respectively, enabling the resonant production of a mono-$Z$ final state discussed below. It can be seen that, in the 2HDM+PS, BR$(H\rightarrow Z a)$ is bigger by roughly a factor of two than BR$(A\rightarrow Z S_2)$ in the 2HDM+S. This is due to the smaller decay width of scalars to quarks compared to pseudoscalars, and therefore a smaller total width, resulting in a larger BR for the 2HDM+S than the 2HDM+PS to the mono-$Z$ final state.
In the 2HDM+S the heavy scalar $S_1$ has a BR of $\mathcal{O}(10\,\%)$ to $\chi\bar\chi$ (from mixing with $S_2$) and $hS_2$, the latter declining with $M_{S_2}$. In the 2HDM+PS the analogue is the heavy pseudoscalar $A$, where BR$(A\rightarrow ah)$ and BR$(A\rightarrow \chi\bar{\chi})$ are of similar strength. The former decay enables a resonant production of the mono-$h$ final state.

\section{Overview of Model Constraints}
\label{sec:phenoconstr}

In this section we give a brief overview on various constraints on the DM properties of the 2HDM+S/PS mainly from non-collider experiments and on 2HDMs in general as they apply to the studied case and reduce the parameter space.

\subsection{Direct Detection}
\label{sec:direct}

Direct detection (DD) phenomenology and their detection prospects are rather different for the two models. In the case of 2HDM+S model the DM is coupled to scalar mediators. It consequently generates Spin Independent (SI) interactions due to t-channel exchange of the scalar mediators at tree level, described by the following operator~\citep{Bell:2016ekl,Bell:2017rgi}:
\begin{equation}
    \mathcal{L}_N=c_N\, \mathcal{O}_1^N= c_N\, \bar\chi\chi\, \bar N N
\end{equation}
where: 
\begin{equation}
    c_N=m_N\ \frac{y_\chi \sin\theta \cos\theta}{v} \left(\frac{1}{M_{S_1}^2}-\frac{1}{M_{S_2}^2}\right)\left(\epsilon_u f_{T_u}^N +\epsilon_d \sum_{q=d,s}f_{T_q}^N+\frac{2}{9} \frac{2 \epsilon_u+\epsilon_d}{3}\ f_{T_g}^N\right)
\end{equation}
with the coefficients $\epsilon_q$ as given in Tab.~\ref{tab:coeffs} and the coefficients $f$ can be found in \cite{DelNobile:2013sia}. Such interaction is well within the reach of current detectors for $y_\chi\sim \mathcal{O}(1)$, $\sin\theta\cos\theta \sim \mathcal{O}(0.3)$, $\tan\beta\sim\mathcal{O}(1)$ and $M_{S_i}\sim\mathcal{O}(\TeV)$, apart from the regions where there are negative interference effects \citep{Bell:2016ekl}. In the following sections on collider searches we focus on light DM where DD is insensitive, therefore acting as a complementary search.

In the case of the 2HDM+PS model, the DM-nucleon interaction, from tree-level exchange of pseudoscalar mediators is described by the following operator~\cite{DelNobile:2013sia}:
\begin{equation}
    \mathcal{L}_N=\tilde{c}_N \mathcal{O}_4^N=c_N \bar \chi i \gamma^5 \chi \, \bar N i \gamma^5 N,
\end{equation}
where the effective coefficient $\tilde{c}_N$ is given, for example, in~\cite{Dolan:2014ska,Arina:2014yna}. This operator, in the non-relativistic limit, reduces to:
\begin{equation}
    \mathcal{L}_N \equiv 4 \tilde{c}_N (\vec{s}_\chi \cdot \vec{q}) (\vec{s}_N \cdot \vec{q}),
\end{equation}
where $\vec{s}_\chi,\vec{s}_N$ are the DM and nucleon spins, respectively, while $\vec{q}$ is the momentum transfer. The corresponding cross-section is hence suppressed by the fourth power of the momentum transfer and experimentally relevant only for very light masses of the mediators~\cite{Dolan:2014ska,Arina:2014yna,Sanderson:2018lmj}. Spin independent DM nucleon interactions, within the reach of, especially, next generation detectors like XENONnT and DARWIN, emerge however at the one loop level~\citep{Arcadi:2017wqi,Sanderson:2018lmj,Abe:2018emu,Ertas:2019dew,Abe:2019wjw} (see also~\cite{Freytsis:2010ne,Ipek:2014gua}).

\subsection{Indirect Detection and Relic Density}
\label{sec:indirect}

Adopting the WIMP paradigm, the DM relic density, as well as eventual Indirect Detection (ID) signals, rely on DM annihilation processes. 

The case of the 2HDM+S model has been extensively studied e.g.~in \citep{Bell:2017rgi}. The possible annihilation channels include SM fermion-antifermion pairs as well as two scalars and scalar-gauge bosons final states. The full list of annihilation channels is $f\bar f$, $S_1 S_1$, $S_1 S_2$, $S_2 S_2$, $H^+ H^-$, $H^{\pm}W^{\mp}$, $AA$, $AZ$, $S_1 h$, $S_2 h$, where $f$ denotes all SM fermions with $m_f<m_\chi$. All these annihilation channels are characterized by p-wave suppressed annihilation cross-sections. While it is possible to achieve the correct relic density by suitably accommodating the model parameters, we do not expect signals measurable through ID experiments.

The phenomenology related to ID and relic density for the 2HDM+PS model has been extensively reviewed in \cite{Abe:2018bpoWG} (see also \cite{Ipek:2014gua,Arcadi:2017wqi,Arcadi:2019lka}).
The DM relic density is determined by annihilation processes into $f\bar f$, $hA$, $HA$, $H^{\pm}W^{\mp}$, $HZ$, $aa$, $aA$, and $AA$ final states. The first two types of annihilation channels have s-wave dominated cross-sections, hence also relevant for ID (see below). The annihilation cross-sections into pseudoscalars are, instead, p-wave suppressed. Nevertheless, the $aa$ channel can be relevant for the relic density, especially for very light $a$.

Concerning ID, possible signals are mostly accounted by the $b\bar b$, $t\bar t$ and $ha$ channels, the latter giving a four fermion, typically $b \bar b b \bar b$, signature. For heavy DM, the $hA$, $HZ$, and $H^{\pm}W^{\mp}$ final states might play a role as well. The $aa$, $aA$, and $AA$ final states contribute to ID to a negligible extent, because the corresponding cross-sections are p-wave suppressed. As discussed in~\cite{Abe:2018bpoWG}, FERMI-LAT bounds on the $b\bar b$ and $t\bar t$ channels~\cite{Fermi-LAT:2016uux} can probe DM masses from $190\GeV$ up to around $400\GeV$, therefore offering a complementary approach to the collider searches. No dedicated studies exist, instead, for annihilation into a gauge and a Higgs boson.

\subsection{Flavour Constraints}
\label{sec:flavour}

As already mentioned, we are considering specific realizations of the 2HDM in which flavor violation processes are forbidden at tree level through suitable combinations of the couplings of the SM fermions with the different Higgs doublets. FCNC processes can be nevertheless induced at loop level. A comprehensive discussion of the possible bounds has been performed in \cite{Enomoto:2015wbn}. Among them, the most relevant ones come from $b \to s$ transitions, in particular the $B \to X_s \gamma$ process, whose rate is mostly sensitive to $\tan\beta$ and $M_{H^{\pm}}$. $b \rightarrow s$ transitions mostly constrain 2HDM realizations featuring $\tan\beta$ enhanced interactions of the BSM scalars with $d$-type quarks, namely the type II and type Y models. For them we have a lower bound $M_{H^{\pm}}>570\,\mbox{GeV}$ weakly sensitive to the value of $\tan\beta$. The bound on $M_{H^{\pm}}$ can be translated into bounds for the masses of the other extra scalars in view of the relations imposed by bounds from EWPT and the scalar potential, as illustrated in the subsections below (see also e.g. \cite{Arcadi:2018pfo,Arcadi:2019lka}). Additional constraints, from $B_s \rightarrow \mu^+ \mu^-$ and $B \rightarrow K\mu^+ \mu^-$ processes, affect the type~II model for moderate/high values of $\tan\beta$. Models of type~I and type~X , having $\tan\beta$ suppressed interactions with the SM quarks, are constrained only in a small region of the parameter space for $\tan\beta \lesssim 2$.

\subsection{Electroweak Precision Constraints}
\label{sec:ewprecision}

It can be shown that the scalar potential for the 2HDM+S in Eq.~\eqref{eq:fullpotential} and~\eqref{eq:potentialsh} breaks the custodial symmetry \citep{Haber:1992py,Pomarol:1993mu,Barbieri:2006dq,Gerard:2007kn,Grzadkowski:2010dj,Haber:2010bw} (in particular the $\lambda_4,\, \lambda_5$, and $\lambda_{hHS}$ terms) and leads to contributions to EW precision observables, unless $M_A=M_{H^+}$. The most relevant observable is the $\rho$ parameter, which receives a contribution of
\begin{align}
    \Delta\rho = \frac{1}{(4\pi)^2\, v^2}\, \big( &F\!\left(M_{H^+}^2,M_A^2 \right) + \cos^2\theta \, F\!\left(M_{H^+}^2,M_{S_1}^2 \right)+\sin^2\theta \, F\!\left(M_{H^+}^2,M_{S_2}^2\right) \nonumber\\
    &\left. -\cos^2\theta \, F\!\left(M_A^2,M_{S_1}^2\right)-\sin^2\theta \, F\!\left(M_A^2,M_{S_2}^2\right) \right), 
\label{eq:rhoEWS}
\end{align}
where
\begin{equation}
    F(x,y) \equiv \frac{x+y}{2}-\frac{xy}{x-y}\, \log(x/y).
\end{equation}
One can verify that $\Delta\rho=0$ for $M_A=M_{H^+}$. Similarly, for the 2HDM+PS potential in Eq.~\eqref{eq:potentialp} and~\eqref{eq:potentialph},
\begin{align}
    \Delta\rho = \frac{1}{(4\pi)^2\, v^2}\, &\left( F\!\left(M_{H^+}^2,M_H^2\right) + \cos^2\theta \, F\!\left(M_{H^+}^2,M_{A}^2\right) +\sin^2\theta \, F\!\left(M_{H^+}^2,M_{a}^2\right) \right.\nonumber\\
    &\left. -\cos^2\theta \, F\!\left(M_A^2,M_H^2\right)-\sin^2\theta \, F\!\left(M_a^2,M_H^2\right) \right), 
\label{eq:rhoEWPS}
\end{align}
and $\Delta\rho=0$ for $M_H=M_{H^+}$. Thanks to our assumption of mass degeneracy, $M_H = M_{H^\pm} = M_A$ or $M_A = M_{H^\pm} = M_{S_1}$, we automatically fulfill EW precision constraints. However, as soon as a small mass splitting appears in the spectrum, limits from EW precision can become very constraining if the scalars making up the doublet do not have very similar masses. This implies $M_{S_1}\sim M_A\sim M_{H^\pm}$ and $0\le\theta\lesssim \pi/4$ for the 2HDM+S and $M_{A}\sim M_H\sim M_{H^\pm}$ and $0\le\theta\lesssim \pi/4$ for the 2HDM+PS. This, together with the fact that to have resonant enhancement of mono-$Z$ and mono-$h$ one of the (pseudo)scalars needs to be below and the other above the top threshold, is the reason that motivates our choice of concentrating on $M_{S_1}>M_{S_2}$, respectively $M_A>M_a$. For more details on EW precision constraints for these models, see \citep{Bell:2017rgi, Bauer:2017ota}.

\subsection{Perturbativity and Unitarity Constraints}
\label{sec:unitarity}

Perturbativity and unitarity constraints for the 2HDM+S model have been studied in detail in \citep{Bell:2016ekl} (see also \citep{Kanemura:2015ska,Klimenko:1984qx}). By requiring that the couplings of the scalar potential are perturbative and that the amplitude of scattering processes of the type $\phi_i \phi_j \rightarrow \phi_k \phi_l$ involving scalar states preserves unitarity, we obtain the following constraints:
\begin{equation}
\begin{aligned}
\label{eq:2HDMS_unitarity}
    & |\lambda_3|+|\lambda_4| <1, \\
    & |\lambda_3|+|\lambda_5| <1 ,\\
    & \lambda_1+\lambda_2+\sqrt{\lambda_1^2-2 \lambda_1 \lambda_2 +\lambda_2^2+4 \lambda_5^2}<2, \\
    & \lambda_1+\lambda_2+\sqrt{\lambda_1^2-2 \lambda_1 \lambda_2 +\lambda_2^2+4 \lambda_4^2}<2, \\
    & \lambda_{11S}+\lambda_{22S}+\sqrt{\lambda_{11S}^2-2 \lambda_{11S}\lambda_{22S} +\lambda_{22S}^2+4 \lambda_{12S}^2}<2 .
\end{aligned}
\end{equation}
These bounds on the quartic couplings can be also re-expressed into upper limits on the mass splittings between the scalar Higgs eigenstates (see, for example, \citep{Bell:2016ekl}). Together with Eq.~\eqref{eq:2HDMS_unitarity} one should also take into account the following constraints from the requirement that the scalar potential is bounded from below:
\begin{align}
    & \lambda_{1,2,S}>0, \nonumber\\
    & \sqrt{\lambda_1 \lambda_2}>-\lambda_3,\nonumber\\
    & \sqrt{2 \lambda_{1}\lambda_S}>-\lambda_{11S,}\\
    & \sqrt{2 \lambda_{2}\lambda_S}>-\lambda_{22S},\nonumber\\
    & \sqrt{\lambda_1 \lambda_2}>|\lambda_5|-\lambda_3-\lambda_4.\nonumber
\end{align}
Unitarity constraints for the 2HDM+PS model have been considered, instead, in \citep{Goncalves:2016iyg}. These rely on the amplitudes of the processes $aa$, $aA$, $AA \to WW$ and can be expressed through the following condition:
\begin{equation}
    |\Lambda_{\pm}|< 8 \pi,
\end{equation}
where
\begin{equation}
    \Lambda_{\pm}=\left[\frac{\Delta_H^2}{v^2}-\frac{\Delta_a^2 \left(1 - \cos4 \theta\right)}{8v^2}\pm \sqrt{\frac{\Delta_H^4}{v^4}+\frac{\Delta_a^4 \left(1 - \cos4 \theta\right)}{8v^4}}\right],
\end{equation}
with $\Delta_a^2=M_A^2-M_a^2$ and $\Delta_H^2 = \frac{-2 \hat{M}_{12}^2}{\sin 2 \beta} - M_{H^{\pm}}^2 + 2 M_W^2 - \frac{M_h^2}{2}$.
Requiring perturbativity on top of unitarity strengthens the limit down to
\begin{equation}
    |\Lambda_{\pm}|< \mathcal{O}(1).
\end{equation}

\subsection{Collider Searches}
\label{sec:collider}

In this section, we will focus on the collider searches for $t \bar{t} + \met$ and mono-jet, whereas $t \bar{t}$, mono-$Z$ and mono-$h$ will be discussed in detail in Sec.~\ref{sec:model_comparison}, since they lead to the most stringent limits for the 2HDM+S/PS.

\subsubsection{$t\bar{t}+\met$}

New spin-0 mediators with large invisible widths can be searched in the $t\bar{t}+\met$ (and $b\bar{b}+\met$) channels. The most recent experimental searches have been reported in \citep{Aaboud:2017rzf,Sirunyan:2019gfm}. Their results have been interpreted in the context of the so-called DM-simplified model (DMF). They can be, however, straightforwardly applied to the 2HDM+PS model, as long as $M_a \ll M_A$, by applying the simple scaling relation (a similar relation applies for $b\bar{b}+\met$  with the replacement of $\tan\beta$ according to Tab.~\ref{tab:coeffs})
\begin{equation}
\label{eq:simple_rescaling}
    \frac{\sigma(pp \to t\bar{t}+\met)_{2HDM+PS}}{\sigma(pp \to t\bar{t}+\met)_{DMF}}={\left(\frac{y_\chi \sin\theta}{g_\chi g_q \tan\beta}\right)}^2.
\end{equation}
As discussed in \citep{Abe:2018bpoWG}, the limits of \citep{Aaboud:2017rzf,Sirunyan:2019gfm} can be applied, in analogous manner as illustrated above, also to the 2HDM+S, as long as there is substantial mass splitting between the BSM spin-0 scalars. 

In case that the new spin-0 scalars have comparable masses, the $\met$ spectrum features distortions with respect to the DMF case. More refined procedures to map the experimental limits on the models under study should then be applied. The case of the 2HDM+PS model is illustrated in~\citep{Abe:2018bpoWG}.

For the extended 2HDMs we are considering, the $t\bar{t}+\met$ exclusions are shown to be subdominant~\cite{Aaboud:2019yqu}. Therefore, we don't derive explicit bounds from those searches.

\subsubsection{Mono-jet}

Similarly to what occurs in heavy flavors + missing energy searches, observational constraints are typically interpreted in the context of simplified models. In the limits in which the 2HDM-like neutral bosons are decoupled, in mass, with respect to the singlet-like states, the kinematic distribution of the mono-jet events are substantially the same for the DMF and 2HDM+S/PS models. Experimental limits can be then applied to our setups just by using analogous scaling relations to Eq.~\eqref{eq:simple_rescaling}. 

As mono-jet events provide the strongest bounds among the initial state radiation signatures~\cite{Bernreuther:2018nat}, we explicitly checked promising points with the help of the CheckMATE~\cite{Dercks:2016npn} implementation of the latest ATLAS search~\cite{Aaboud:2017phn} and found no excluded points for $\tan\beta =1$. As shown in~\cite{Bauer:2017ota}, mono-jet limits only arise for $\tan\beta < 1$, however, this region is already excluded by other constraints and therefore we do not investigate this further.

\section{Comparison of 2HDM + Scalar/Pseudoscalar and their LHC Signatures}
\label{sec:model_comparison}

After the general overview of constraints in the last section, we will now turn to the 2HDM+S/PS specific LHC signatures and limits we obtained for the leftover parameter space by dedicated collider simulations.

\subsection{General Aspects}
\label{sec:general_aspects}

Before describing the simulations and discussing our parameter space in detail, let us first have a brief look at resonant production processes, which can be described analytically and will become handy for the interpretation of the simulation results.

\paragraph{Resonant production}
The production cross section for a spin-0 scalar (or pseudoscalar) resonance $S$, with mass $M$ and total decay width $\Gamma_{\rm tot}$, subsequently decay into the final state $X$, can be schematically written, in the narrow width approximation, as \cite{DEramo:2016aee}
\begin{equation}
    \sigma(p p \to S \to X) = \frac{\Gamma(S\to X)}{M \Gamma_{\rm tot} s} \sum_{i} C_{i} \Gamma(S\to i) = \frac{1}{M s}\, {\rm BR}(S\to X) \sum_{i} C_{i} \Gamma(S\to i) .
\label{eq:sigmaresonant}
\end{equation}
The sum over $i$ runs over the possible partonic initial states, for example quark or gluon pairs, $C_i$ are weight factors that account for the protons PDFs and colour factors, and $s=(13\,\mathrm{TeV})^2$ is the center of mass energy squared. The values of the $C_i$ are given by:
\begin{align}
    C_{gg} &= \frac{\pi^2}{8} \int_{M^2/s}^1 \frac{\mathrm{d}x}{x} g(x) \,g\!\left(\frac{M^2}{sx}\right),\\
    C_{q\bar{q}} &= \frac{4\pi^2}{9} \int_{M^2/s}^1 \frac{\mathrm{d}x}{x} \left[q(x)\,\bar{q}\!\left(\frac{M^2}{sx}\right) + q\!\left(\frac{M^2}{sx}\right)\bar{q}(x)\right],
\end{align}
with $q(x)\left(\bar{q}(x)\right)$ and $g(x)$ being, respectively, the parton distribution functions (PDFs) of quarks (anti-quarks) and gluons inside the proton and $x$ is the conventional Bjorken scaling variable.

\begin{figure}[t]
\begin{center} \hspace{-0.4cm}
    \includegraphics[width=0.98\linewidth]{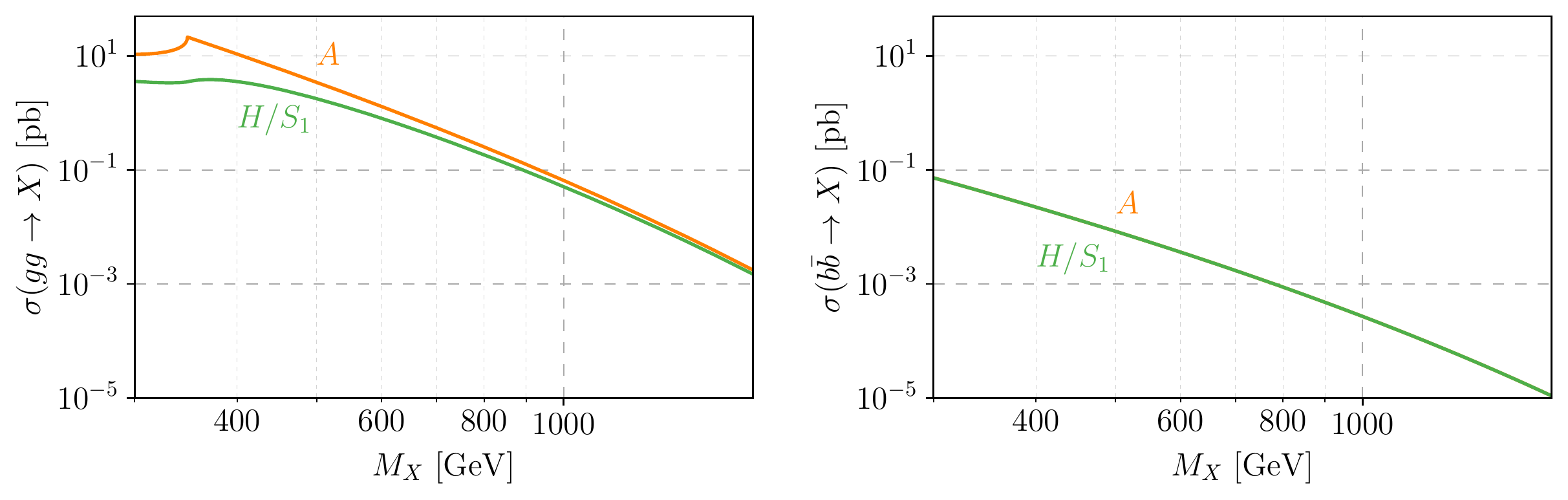} 
    \caption{Production cross sections for the heavy scalar $H / S_1$ and pseudoscalar $A$ through gluon fusion (left) and $b\bar{b}$ initial state (right), in the 2HDM+PS/S as given in Eq.~\eqref{eq:sigmaresonant}. The plot depicts the case of $\tan\beta=1$ and $\sin\theta=0$.}
    \label{fig:monozsigma}
\end{center}
\end{figure}

In general, the two dominant production mechanisms are gluon fusion and $b\bar{b}$ initial states. We show the production cross sections of the heavy (pseudo)scalar, as a function of the resonance mass for both production modes in Fig.~\ref{fig:monozsigma}. The left panel shows the production cross section for gluon fusion, the right panel for $b\bar{b}$ initial states. For $\tan \beta = 1$ and $\sin\theta=0$ the production cross section from gluon fusion results to be about 100 times larger than the one from $b\bar{b}$ initial states. This dominance of gluon fusion production depends on the 2HDM Yukawa type in combination with the value of $\tan\beta$, but holds for all 2HDM Yukawa types in the whole parameter space we take into consideration ($0.3 \leq \tan \beta \leq 3$). Therefore, in our simulations we can safely neglect the $b\bar{b}$ production mode.

From Fig.~\ref{fig:monozsigma}, we can also see that the production of a pseudoscalar particle always results in a larger cross section. This is because the effective coupling of the pseudoscalar mediator is larger, from same assignation of the Yukawa couplings, with respect to the scalar mediator, see e.g.~\cite{Gunion:1989we,Arcadi:2019lka}.

\paragraph{Parameter overview}

For a better overview we briefly summarize our choice of parameter values, which are motivated by various constraints discussed above and used for the following Monte Carlo simulations and other examples throughout the paper:
\begin{align} \label{eq:parameters}
    & M_A = M_{H^\pm} = M_{H/S_1}\nonumber\\
    & \lambda_3 = M_h^2/v^2 \nonumber\\
    & \lambda_{iiS/P} = 0\nonumber\\
    & y_\chi^{S,PS} = 1\\
    & m_\chi = 10\, \text{GeV}\nonumber\\
    & \sin\theta = 0.3\nonumber\\
    & \tan\beta \in [0.3, 3] . \nonumber
\end{align}

\paragraph{Signal Generation}
\label{sec:signal}

We simulate events for the signal processes with MadGraph5\_aMC@NLO~\cite{Alwall:2014hca, Hirschi:2015iia, Ossola:2007ax, Peraro:2014cba, Denner:2016kdg} at next-to-leading order (NLO) in QCD using the 263000 PDF set (NNPDF3.0)~\cite{Butterworth:2015oua} provided through LHAPDF6~\cite{Buckley:2014ana}.
For the parton-showering we use the MadGraph built-in Pythia~$8.2$~\cite{Sjostrand:2014zea}. A fast detector simulation is done with Delphes~$3.4.2$~\cite{deFavereau:2013fsa} using the provided CMS detector card.
The final cuts are implemented in MadAnalysis~$5$~\cite{Conte:2018vmg, Conte:2014zja}.
The correct implementation of the program chain and analysis is checked by reproducing the mono-$Z$ and mono-$h$ exclusions for the 2HDM+PS presented in \cite{Abe:2018bpoWG, Aaboud:2019yqu}.

\par\vspace{\baselineskip}\noindent
In the following, we will look at the different channels that provide the strongest limits for the 2HDM+S/PS model.

\subsection{$t\bar t$ Resonances}
\label{sec:di-top}

For masses above the top-threshold, $M_{H/S_1,\, A} > 2\,m_t$, the additional heavy Higgses dominantly decay into a pair of top quarks, cf.~Sec.~\ref{sec:decays}. Searches for $t\bar t$ resonances are therefore a powerful tool to test 2HDMs in general and 2HDM+S/PS in particular.

One aspect that complicates the analysis, is that the signal processes interfere non-trivially with the SM background, as pointed out and taken into account by the experimental analysis in~\cite{Aaboud:2017hnm}. There the ATLAS results are interpreted in a 2HDM of type II, for the case in which the individual contribution of the two mediators to the signal can be singled out\footnote{Physically, this could for example correspond to the limit where one mediator is much heavier than the other.}
as well as for the mass-degenerate case where both mediators contribute. As the later case gives significantly stronger constraints, deriving exclusions for a single mediator serves as a conservative estimate.

The most recent bounds for $\sqrt{s}=13$\,TeV and an integrated luminosity of $35.9$\,fb$^{-1}$ are provided by CMS in~\cite{Sirunyan:2019wph}. The limits are presented in terms of simplified models with either a scalar or a pseudoscalar and Yukawa-like couplings to tops and in the hMSSM. Here, a $1.9\,\sigma$ ``signal-like deviation'' is observed that would fit to a pseudoscalar with a mass of around $400$\,GeV, therefore the limits in that mass range are not significantly stronger than the ones from the previous ATLAS analysis with $\sqrt{s}=8$\,TeV.

The additional light state $a/S_2$ can have a non-trivial impact on the limit due to interference effects, cf.~Sec.~$7.1$ in~\cite{Abe:2018bpoWG}. However, this effect is expected to be small for our choice of small mixing $\sin\theta=0.3$. A detailed analysis of the impact of interference and combining the limits for the two heavy states is beyond the scope of this article. Instead, we recast the recent CMS limits~\cite{Sirunyan:2019wph} for single mediators to our parameter space, interpolating between the different total width to mass ratios given in the paper. As commented above in the context of~\cite{Aaboud:2017hnm}, this can bee seen as a conservative estimate, since the limits get significantly stronger by taking the contribution of both mass-degenerate states $H/S_1$ and $A$ into account.

\begin{figure}[t]\begin{center}
    \includegraphics[width=0.8\linewidth]{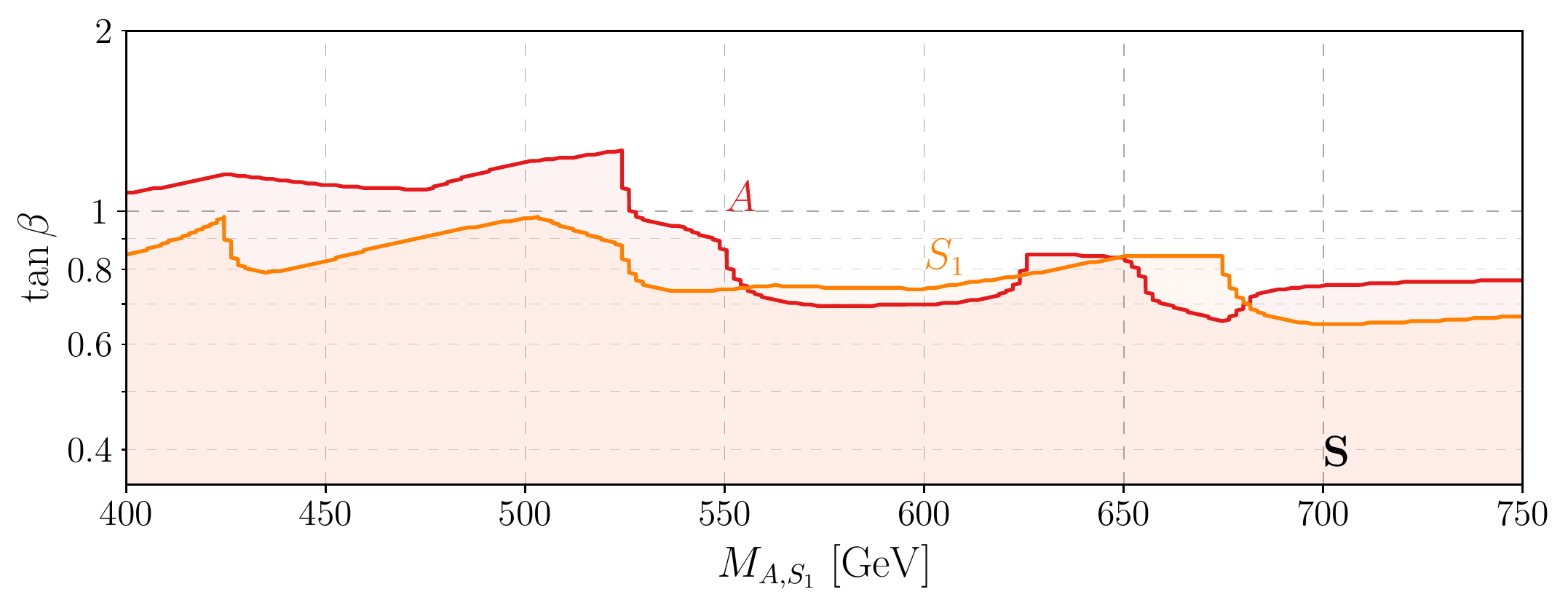}\\
    \includegraphics[width=0.8\linewidth]{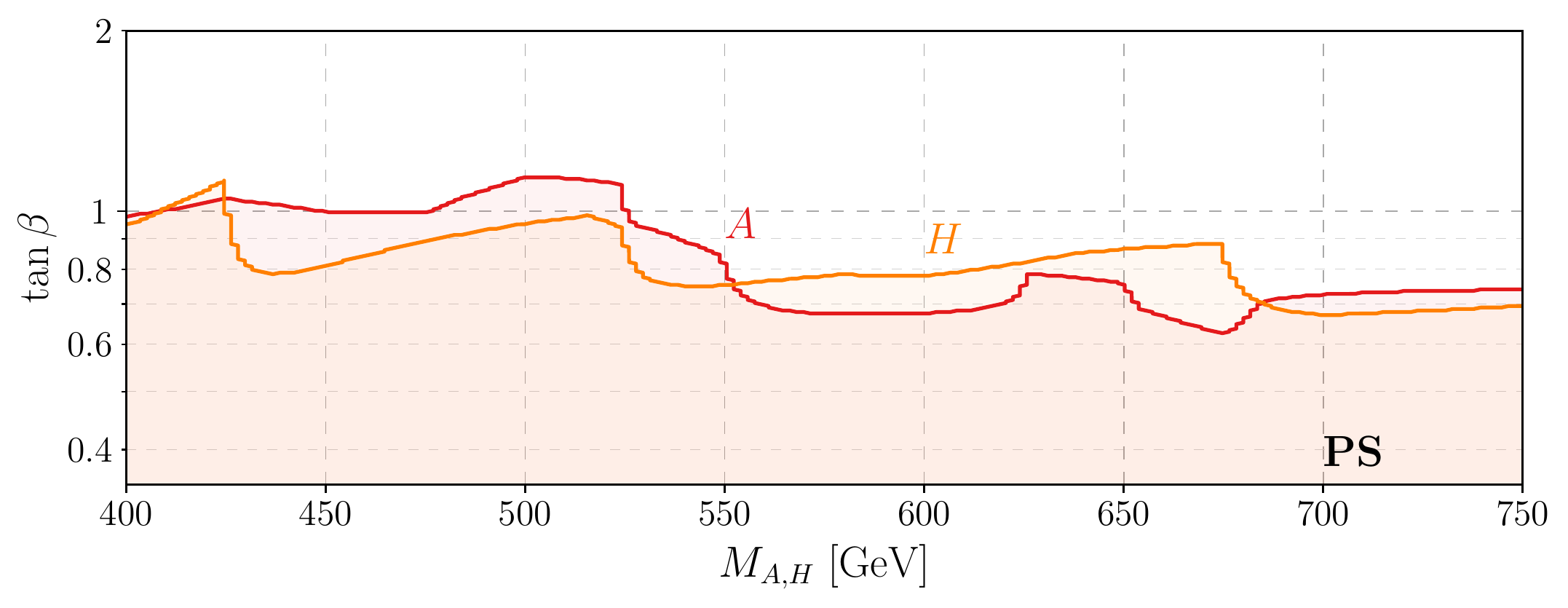}
    \caption{The $2\,\sigma$ observed exclusion limits from $t\bar t$ resonance searches by CMS~\cite{Sirunyan:2019gfm} in the $M_{A,H/S_1}$--$\tan\beta$--plane for the 2HDM+S (top) and 2HDM+PS (bottom). The parameters are fixed to the values from Eq.~\eqref{eq:parameters} with $M_{a/S_2} = 400$\,GeV and loosening the assumption of mass-degeneracy. The limits derived from searches for a scalar (pseudoscalar) resonance $H/S_1$ ($A$) are given in orange (red).}
\label{fig:di-top_limits}
\end{center}
\end{figure}

The limits are shown in Fig.~\ref{fig:di-top_limits} in the $M_{A,H/S_1}$--$\tan\beta$--plane for the 2HDM+S (top) and 2HDM+PS (bottom). We choose the setting in Eq.~\eqref{eq:parameters} with $M_{a/S_2}=400$\,GeV and loosen the assumption of mass-degeneracy. There exists a mild dependency on the mass of the light state due to changes of the total width which is shown in Fig.~\ref{fig:mMcomb} and \ref{fig:mtbetacomb}.

The decay widths of pseudoscalars to quarks and gluons (for $M_A$ above the top threshold) are bigger than the ones of scalars, cf.~Sec.~\ref{sec:decays}, App.~\ref{sec:Gamma_app} and \cite{Djouadi:1995gv}. Together with the effective coupling approximation for the dominant gluon fusion production, cf.~Eq.~\eqref{eq:sigmaresonant} and~\cite{Hahn:2006my}, it can be understood that pseudoscalar resonances are expected to provide a stronger constraint than scalar ones.
In the 2HDM+S the couplings of the heavy scalar $S_1$ are reduced due to the mixing with the singlet $S_2$, while the pseudoscalar couplings are untouched. As a consequence the exclusion for $S_1$ reaches up to $\tan\beta=1$, while the one for $A$ clearly exceeds $\tan\beta=1$.
In the 2HDM+PS instead, the stronger constrained pseudoscalar mixes with the singlet $a$, weakening the exclusion and leading to overall weaker constraints from $t\bar t$ searches for this model.
In both models, for masses of the heavy Higgs around $500$\,GeV, $t\bar t$ resonance searches provide a strong lower limit on $\tan \beta$ and essentially exclude values below $\tan \beta \lesssim 1$.

\subsection{Mono-$Z$}
\label{sec:missz}

\begin{figure} \centering 
    \includegraphics[width=0.8\linewidth]{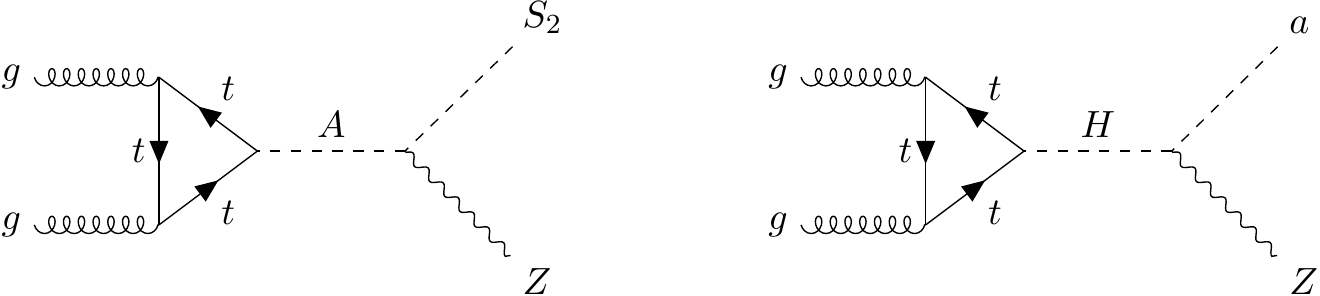}
    \caption{Feynman diagrams for the resonant production of a mono-$Z$ signal via gluon fusion production in the chosen mass hierarchy for the 2HDM+S (left) and 2HDM+PS (right).}
\label{fig:monozfeyn}
\end{figure}

For a strong mono-$Z$ signal, which is a general feature for 2HDMs extended by pseudoscalar mediated DM~\cite{Bauer:2017fsw}, the heavy neutral spin-0 particle of the doublet which doesn't mix with the singlet has to be produced, meaning the scalar $H$ in the 2HDM+PS and the pseudoscalar $A$ in the 2HDM+S. This resonantly produced state can decay to a $Z$ boson and the light state $a$ or $S_2$, respectively, which further decay to $\chi\bar\chi$ with a high branching ratio, cf.~Sec.~\ref{sec:decays}. The Feynman diagrams for the processes are shown in Fig.~\ref{fig:monozfeyn} for the dominant gluon fusion production. 

Searches for the relatively clean final state, where the $Z$ boson decays leptonically (meaning to electrons or muons), in association with $\met$ are performed by the ATLAS and CMS collaborations~\cite{Aaboud:2017bja, Sirunyan:2017qfc}.

\paragraph{$\met$ Spectrum}
The maximum value of the missing transverse energy $\met$ can be obtained from kinematics and is given by~\citep{Bauer:2017ota} 
\be
    \met^{\mathrm max} = \frac{\lambda^{1/2}(M_{A,H}, M_{S_2,a}, M_Z)}{2M_{A,H}} ,
\ee
where $\lambda(m_1, m_2, m_3)$ is given in Eq.~\eqref{eq:lambda} and the first (second) subscript is used for the 2HDM+S (PS). The missing energy spectrum is given by
\be
    \frac{1}{N}\frac{dN}{d\met} =\frac{\met}{2 \met^{\mathrm max} \sqrt{\left(\met^{\mathrm max}\right)^2 -\met^2}},
\ee
that is a monotonic increasing function in $\met$. However, detector smearing effects are expected to increase the maximum value of $\met$ to $\met^{\mathrm max, D} > \met^{\mathrm max}$ and one expects the resulting distribution to be peaked (instead of having its endpoint) close to $\met=\met^{\mathrm max}$.

\begin{figure}[t] \centering
    \includegraphics[width=0.8\linewidth]{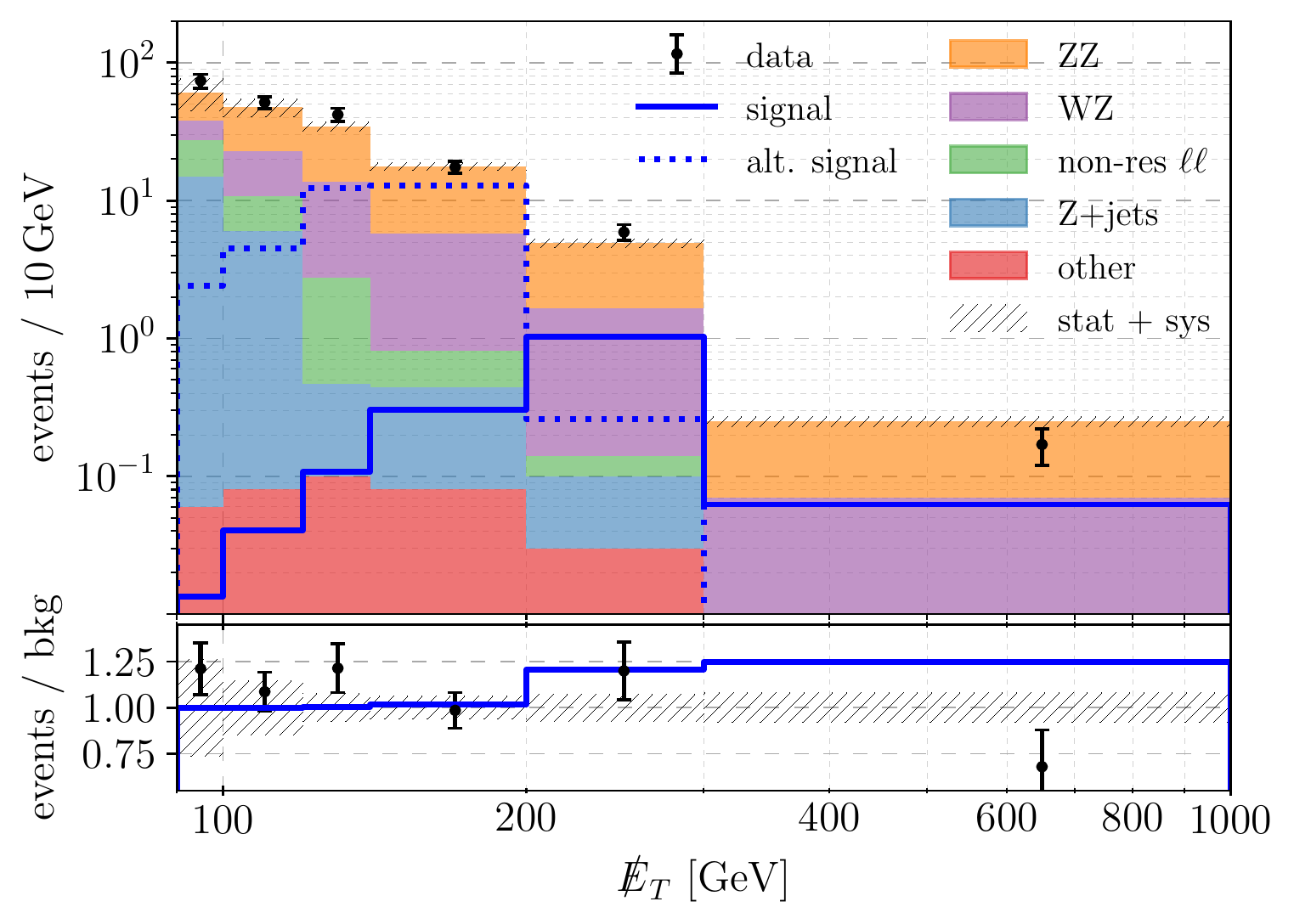}
    \caption{Top panel: $\met$ spectra for $g g \to e^+ e^- \chi \bar{\chi}$ in the 2HDM+PS for $M_a = 250$\,GeV, $M_A = 700$\,GeV (solid line) and $M_a = 150$\,GeV, $M_A = 400$\,GeV (dotted line), while the other parameters are set to the values given in Eq.~\eqref{eq:parameters}. The expected SM backgrounds and observed events are taken from~\cite{Aaboud:2017bja} and shown in different colors with their combined uncertainty displayed as a hatched region on top.\\
    Lower panel: ratios of the observed events (points with error-bars) and signal + background expectation (solid line) to the background expectation plotted together with the background uncertainty (hatched region).}
\label{fig:ETmiss_spectrum}
\end{figure}

Two example $\met$ spectra after detector simulation for the 2HDM+PS in the $e^+e^-+\met$ final state are shown in Fig.~\ref{fig:ETmiss_spectrum} together with the predicted SM backgrounds and the observed number of events as provided by ATLAS~\cite{Aaboud:2017bja}. Nearly identical $\met$ spectra exist for the $\mu^+\mu^-+\met$ final state. As described below both final states are combined to determine the exclusion limits. The signal features a peak a bit below $\met = M_A/2$ on top of the smoothly falling SM background.

\paragraph{Backgrounds}
For mono-$Z$ searches the main irreducible background is $ZZ$ production with one $Z$ decaying to neutrinos. Another important background is $WZ$ production, where one lepton from the $W$-decay escapes detection or a $\tau$ decays hadronically, cf.~Fig.~\ref{fig:ETmiss_spectrum}. Minor backgrounds are $Z+$jets processes with poor $\met$ reconstruction and non-resonant $\ell\ell$ production. In \cite{Aaboud:2017bja}, the backgrounds are estimated from simulations and data-driven methods. For both the electron and muon final state the background uncertainty is dominated by systematic errors. Those are driven by the uncertainties on the $Z+$jets and $ZZ$ in gluon fusion production processes.

\paragraph{Current Constraints}
To determine model constraints we use the ATLAS results~\cite{Aaboud:2017bja} as they are also used by the LHC-DMWG and easier to reproduce than the (slightly stronger) CMS results~\cite{Sirunyan:2017qfc}. To be explicit, for our exclusion bounds, we use the expected number of background events $b$ and the corresponding uncertainty $\sigma_b$ from~\cite{Aaboud:2017bja}, together with our simulated event numbers $s$ for various parameter points and the sensitivity formula from~\cite{Abe:2018bpoWG, Cowan:2012abc}:
\begin{equation}
    Z_i = \sqrt{2 \left( (s + b) \ln\left[ \frac{(s + b)(b + \sigma_b^2)}{b^2 + (s + b) \sigma_b^2} \right] - \frac{b^2}{\sigma_b^2} \ln\left[ 1 + \frac{\sigma_b^2 s}{b\, (b + \sigma_b^2)} \right] \right)}.
\end{equation}
The index $i$ refers to the different bins and the values for $Z_i$ are added up quadratically to find the (square of the) overall sensitivity $Z^2$, where one expects to exclude parameter points with $Z>2$ at $95\,\%$ confidence level. In this way we obtain the expected exclusion limits, however, as they are very similar to the observed limits, cf.~\cite{Aaboud:2019yqu}, we can use the expected mono-$Z$ limits to compare them to the observed $t \bar{t}$ and mono-$h$ ones in Sec.~\ref{sec:combined_constraints}.

\begin{figure}[t] \begin{center}
\hspace{-0.8cm}
    \includegraphics[width=0.98\linewidth]{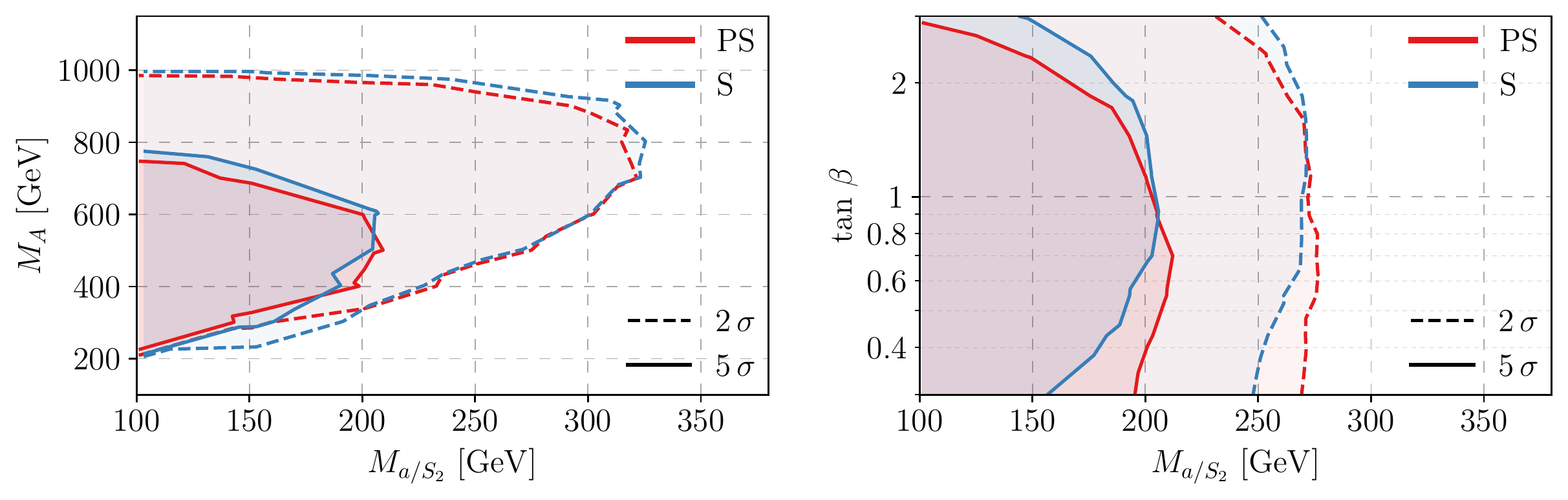}
    \caption{$2\,\sigma$ (dashed) and $5\,\sigma$ (solid) exclusion limits from mono-$Z$ searches in the $M_{a/S_2}$--$M_A$--plane (left) and $M_{a/S_2}$--$\tan\beta$--plane (right) for the 2HDM+S (blue) and PS (red). In both plots we use the parameters as given in Eq.~\eqref{eq:parameters} with $\tan\beta=1$ for the left one and $M_A = 500$\,GeV for the right one.}
\label{fig:monoZ_exc}
\end{center}
\end{figure}

The constraints from mono-$Z$ searches are similar in shape and reach for both models, see Fig.~\ref{fig:monoZ_exc}. This can be understood with help of the approximation in Eq.~\eqref{eq:sigmaresonant} and its corresponding discussion: on the one hand, in the 2HDM+S a heavy pseudoscalar $A$ needs to be produced, which happens with a production cross section that is bigger by roughly a factor of two than the one for the production of the scalar $H$ in the 2HDM+PS, cf. Fig.~\ref{fig:monozfeyn} for the Feynman diagrams.
On the other hand, the relevant branching ratios for the mono-$Z$ final state
\begin{equation}
    \text{BR}(A\to S_2 Z) \approx \text{BR}(H\to a Z)/2 \approx 0.1,
\end{equation}
is bigger in the 2HDM+PS by a factor of approximately two, compensating for the smaller production cross section. The difference in the branching ratios is related to the different decay width of scalars and pseudoscalars into top quarks, as mentioned in Sec.~\ref{sec:decays}.

The general features of the constraints for the mono-$Z$ channel depicted in Fig.~\ref{fig:monoZ_exc} can be readily understood by considering kinematics and the couplings. In the $M_{a/S_2}$--$\tan \beta$--plot, the weakening of the exclusion limit for larger $\tan\beta$ values is due to the fact that the top-coupling scales like $(\tan \beta)^{-1}$ and the top-coupling is essential for the production of the intermediate heavy Higgs, cf. Fig.~\ref{fig:monozfeyn}. For the $M_{a/S_2}$--$M_A$--plot, there are three different features that can be understood: first, the ``diagonal'' lower bound to the exclusion region is due to the fact that for $M_A \lesssim M_{a/S_2} + M_Z$, resonant production is not allowed with on-shell $a/S_2$. Second, the upper bound of the exclusion limit stems from the heavy Higgs being the harder to produce the heavier it is and third, a heavier $M_{a/S_2}$ leaves less energy available for the $Z$, so the $Z$ production gets kinematically suppressed, thereby smoothing out the transition between the first two features.

\subsubsection{Projected Sensitivity}
\label{sec:sensitivity}

\begin{figure}[t]
\begin{center}
\hspace{-0.4cm}
\includegraphics[width=0.98\textwidth]{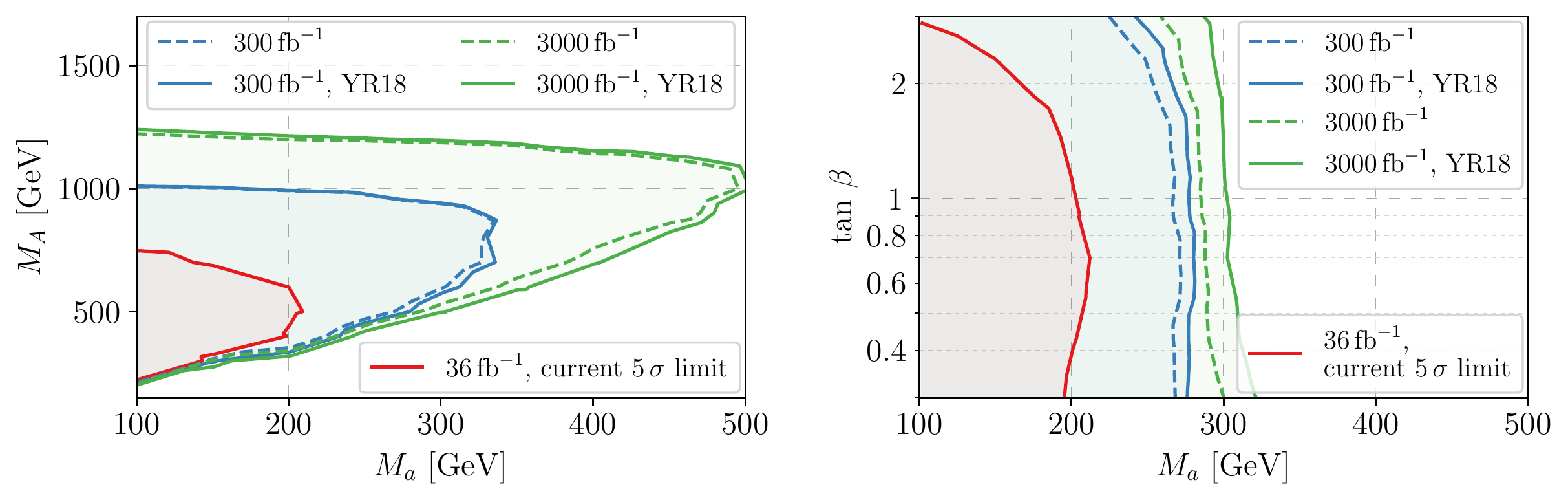}
\caption{Current $5\,\sigma$ exclusion limits for the 2HDM+PS (in red) and projected ones for the high-luminosity LHC (in blue and green) for mono-$Z$ searches in the in the $M_a$--$M_A$--plane (left) and $M_a$--$\tan\beta$--plane (right). The 2HDM+S limits are nearly identical and therefore not shown. The dashed lines correspond to a scenario without any improvement in the systematic uncertainties, whereas the solid lines assume a reduction by $50\,\%$, called YR18 scenario. For the other parameters, the numerical values used are identical to Fig.~\ref{fig:monoZ_exc}.}
\label{fig:monoZ_projection}
\end{center}
\end{figure}

With the detailed data of the experimental analysis at hand~\cite{Aaboud:2017bja}, we estimate the reach of the high luminosity phase of the LHC (HL-LHC). To do so, we assume integrated luminosities of $300$\,fb$^{-1}$ and $3000$\,fb$^{-1}$ and took a projected reduction of systematic uncertainties by $50\,\%$ into account called YR18 scenario~\cite{CMS:2018fux}. The projections for the pseudoscalar model are shown in Fig.~\ref{fig:monoZ_projection} and are nearly identical to the ones for the scalar model, as can be seen for the current bounds in Fig.~\ref{fig:monoZ_exc}. From Fig.~\ref{fig:monoZ_projection}, one can see that the increased integrated luminosity will lead to a substantial increase in sensitivity, strengthening the $5\,\sigma$ limits by roughly a factor of two in terms of the masses for $3000$\,fb$^{-1}$. In contrast to that, the influence of the improved systematic uncertainties will likely be small.

\subsection{Mono-$h$}
\label{sec:missh}

The most recent searches for  mono-$h$ with $h\to b\bar{b}$ by ATLAS and CMS can be found in~\cite{Aaboud:2017yqz, Sirunyan:2018gdw}. Furthermore, in~\cite{Sirunyan:2019zav} CMS performs a first search of mono-$h$ with $h \to b\bar{b},\, \gamma\gamma,\, \tau^+ \tau^-, W^+ W^-,\, ZZ$, where the exclusions are dominated by the $b\bar{b}$ channel. For our analysis, we use the model-independent upper limits on the $h+\met$ cross section provided by ATLAS~\cite{Aaboud:2017yqz} and compare it to the one we find from our simulation, as also done by the LHC-DMWG.

\paragraph{Backgrounds}
The main backgrounds for mono-$h$ with $h \to b\bar b$ searches are $t\bar t$ and vector boson plus multiple jet production, see~\cite{Aaboud:2017yqz, Sirunyan:2018gdw} for a more detailed discussion of the backgrounds and their uncertainty estimates.
Minor contributions arise from single top, diboson, SM  Higgs plus $V (\to\nu\bar\nu)$, and multijet processes.
In the case of the $t\bar t$ background, a bottom quark from a top decay in combination with another $b$-tagged jet can be misidentified as a Higgs boson, while the $\met$ originates from neutrinos.
Similarly, decays of vector bosons also produce $\met$ via neutrinos or missed charged leptons, and jets can mimic a Higgs.
To estimate those backgrounds, control regions with isolated leptons and extended Monte-Carlo simulations are used~\cite{Aaboud:2017yqz, Sirunyan:2018gdw}.

\begin{figure}[t]
\begin{center}
\hspace{-0.8cm}
\includegraphics[width=0.98\linewidth]{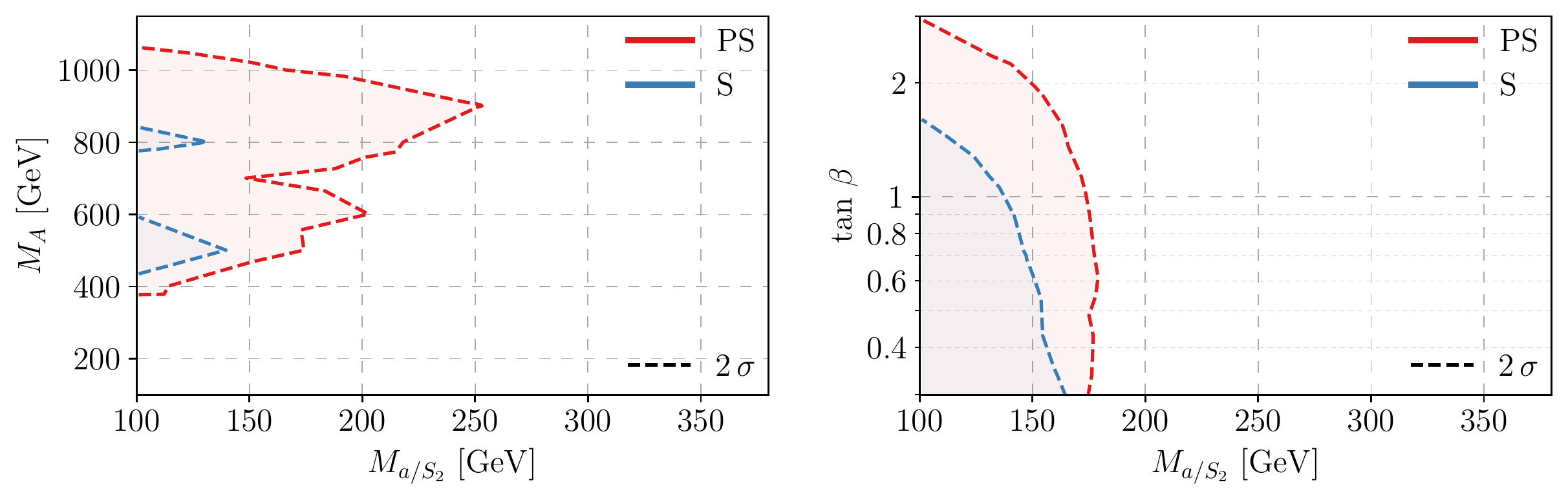}
\caption{$2\,\sigma$ exclusion limits from mono-$h$ searches in $M_{a/S_2}$--$M_A$--plane (left) and  $M_{a/S_2}$--$\tan\beta$--plane (right) for the 2HDM+S (blue) and PS (red). In both plots we use the parameters as given in Eq.~\eqref{eq:parameters} with $\tan\beta=1$ for the left one and $M_A = 500$\,GeV for the right one.}
\label{fig:monoh_exc}
\end{center}
\end{figure}

\paragraph{Current Constraints}
The non-deviation of the observed events from the SM background is translated into (model-independent) upper limits on the $h(b\bar b)+\met$ cross section, which we use in the following by comparing our simulated cross-section to this upper limit, as also done by~\cite{Abe:2018bpoWG}. In~\cite{Aaboud:2017yqz}, only one $\met$ bin is used at a time to minimize the model dependency, which implies that the derived limits are conservative estimates. Furthermore, the dependency of the limit on the kinematics within one bin and on the acceptance and efficiency is calculated for several parameter points of a benchmark model and the least stringent limits are given -- again, this leads to conservative estimates for the exclusion limits of the 2HDM+S/PS, which are shown in Fig.~\ref{fig:monoh_exc}.

Similarly to the mono-$Z$ exclusion plots (cf.~Fig.~\ref{fig:monoZ_exc}), we can see that for larger $\tan\beta$, $M_A$ and $M_{a/S_2}$ the exclusion limits in each case weaken, as they also do in the case of $M_{a/S_2} \sim M_A$. The otherwise most prominent feature in the $M_{a/S_2}$--$M_A$--plane (cf.~left panel of Fig.~\ref{fig:monoh_exc}), is the dip in the exclusion limits around $M_A \sim 700$\,GeV. It originates from a binning effect due to the large $\met$-bins used in the ATLAS analysis. For $M_A>700$\,GeV a significantly higher fraction of the signal events reaches $\met>350$~GeV (as can be seen from the simulated data). Therefore, they end up in the stronger constrained bin reaching from $\met = 350$ to $500$~GeV. This (over-)compensates the decrease in production cross section due to the heavier mediator, which is the dominant effect between $M_A = 600$ and $M_A = 700$\,GeV, thus leading again to stronger limits. The effect of more events ending up in the bin of $\met = 350$ to $500$~GeV can also be understood in the light of the $\met$ spectrum discussion in case of the mono-$Z$ (cf.~Sec.~\ref{sec:missz} and Fig.~\ref{fig:ETmiss_spectrum}). The $\met$ spectrum does not qualitatively change by replacing the $Z$ with an $h$. For values of $M_A>700$\,GeV the peak of the $\met$ spectrum starts to shift from the $\met = 200$ to $350$~GeV bin to the $\met = 350$ to $500$~GeV bin because the peak is located slightly below $\met = M_A / 2$.

\begin{figure} \centering
    \includegraphics[width=0.8\linewidth]{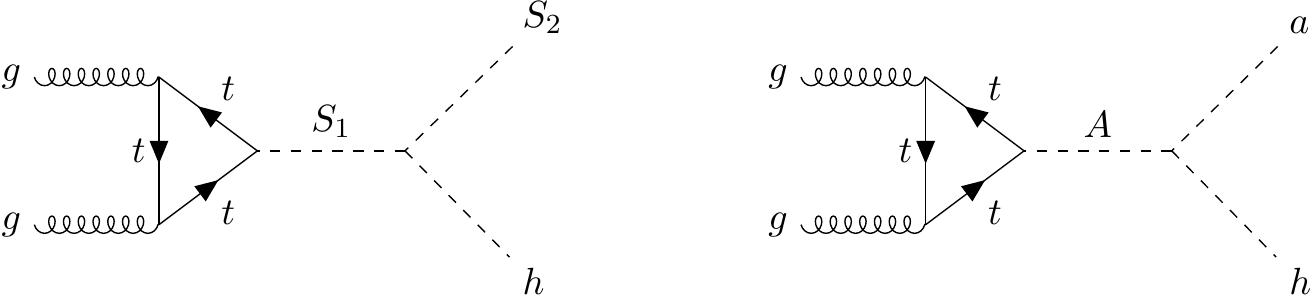}
    \caption{Feynman diagrams for the resonant production of a mono-$h$ signal via gluon fusion production in the chosen mass hierarchy for the 2HDM+S (left) and 2HDM+PS (right).}
\label{fig:monohfeyn}
\end{figure}

In contrast to the mono-$Z$ searches, the exclusions limits from mono-$h$ differ significantly between the 2HDM+S and 2HDM+PS. The upper bound on the mass of the light pseudoscalar $a$ is stronger by approximately a factor two in the 2HDM+PS than the one for the corresponding scalar $S_2$ in the 2HDM+S. This is due to the fact that, to resonantly produce mono-$h$ events in the 2HDM+S, one needs to produce the heavy scalar $S_1$, rather than the the heavy pseudoscalar $A$, and the production cross section for scalars is smaller than the one for pseudoscalars, cf.~Eq.~\eqref{eq:sigmaresonant} and Fig.~\ref{fig:monozsigma}. The opposite happens for the 2HDM+PS, where again the roles of the heavy scalar and pseudoscalar are switched going from mono-$Z$ to mono-$h$, see also the corresponding Feynmann diagrams in Fig.~\ref{fig:monozfeyn} and~\ref{fig:monohfeyn}.
Opposite to the mono-$Z$ case, the branching ratios relevant for mono-$h$ events are similar in both models, cf.~Sec.~\ref{sec:decays}:
\begin{equation}
    \text{BR}(S_1 \to S_2 h) \approx \text{BR}(A \to a h) \approx 0.1 .
\end{equation}
Therefore the 2HDM+PS is expected to have a higher mono-$h$ cross section  in the resonant region and as the kinematics do not change significantly, this results in stronger exclusion bounds.

\subsection{Combined Constraints}
\label{sec:combined_constraints}

\begin{figure}[t]
\begin{center}
\hspace{-0.8cm}
\includegraphics[width=0.98\linewidth]{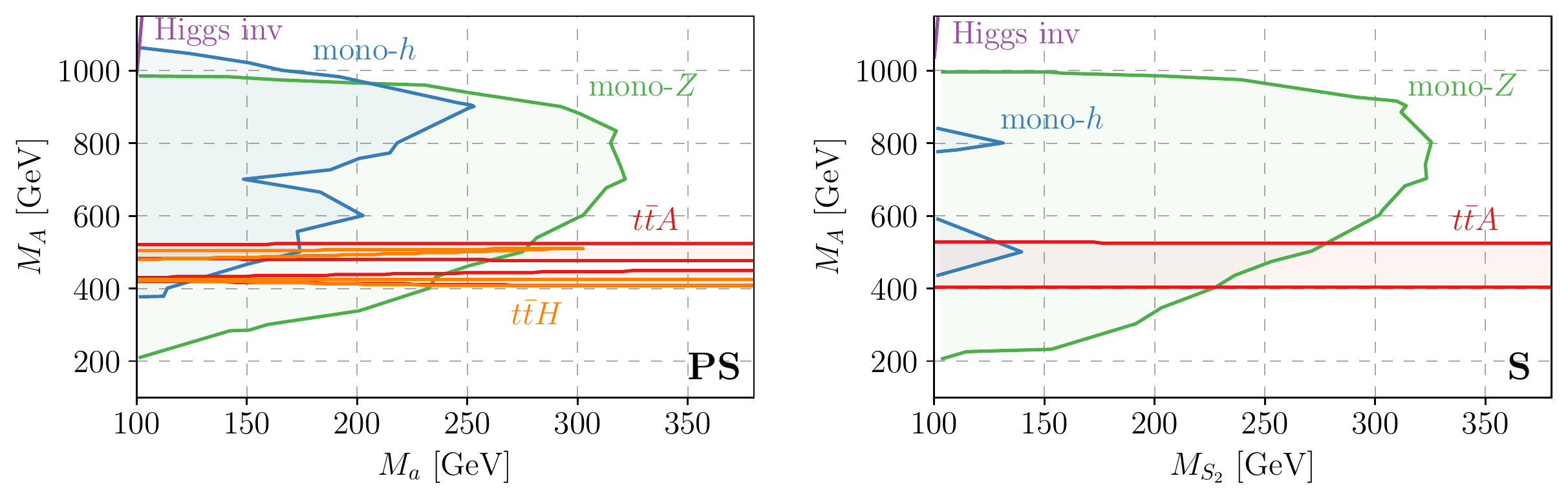}
\caption{Overview of the $2\,\sigma$ exclusion limits from mono-$h$, mono-$Z$, $t\bar t$ resonance (for scalar and psedoscalar mediator) and Higgs to invisible searches (see text for details) in the $M_{a/S_2}$--$M_A$--plane for the 2HDM+PS (left) and 2HDM+S (right). In both plots we use the parameters as given in Eq.~\eqref{eq:parameters} and $\tan\beta=1$.}
\label{fig:mMcomb}
\end{center}
\end{figure}

\begin{figure}[t]
\begin{center}
\hspace{-0.8cm}
\includegraphics[width=0.98\linewidth]{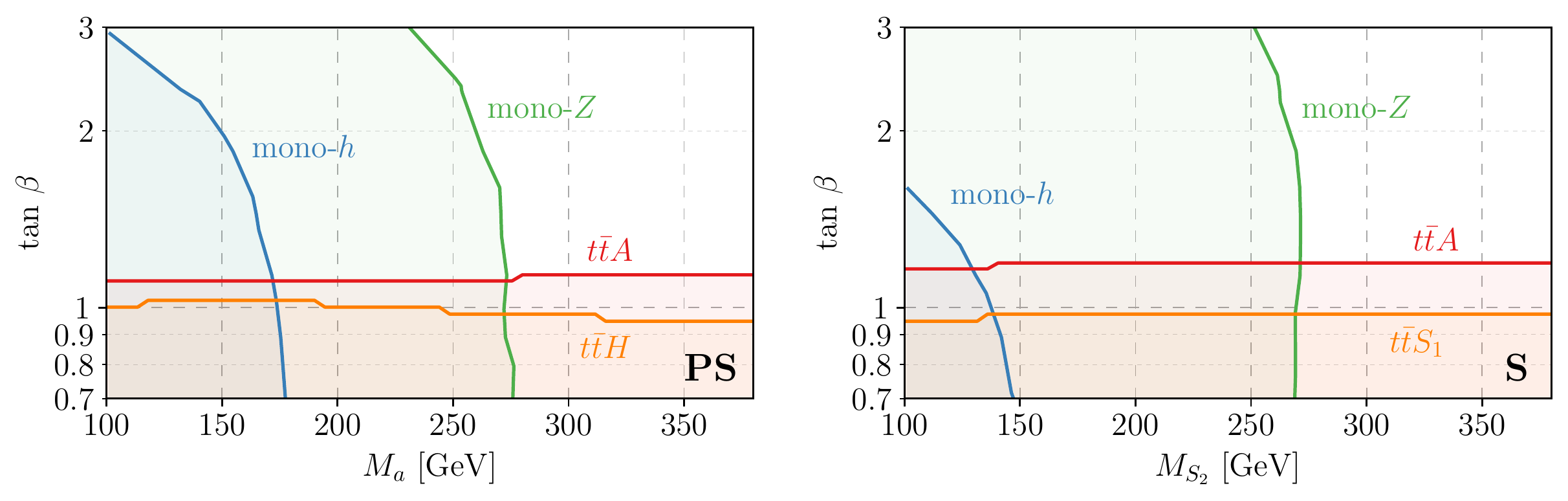}
\caption{Overview on the $2\,\sigma$ exclusion limits from mono-$h$, mono-$Z$, $t\bar t$ resonance (for scalar and pseudoscalar mediator) and Higgs to invisible searches (see text for details) in the $M_{a/S_2}$--$\tan\beta$--plane for the 2HDM+PS (left) and 2HDM+S (right). In both plots we use the parameters as given in Eq.~\eqref{eq:parameters} and $M_{H/S_1} = M_{H^\pm} = M_A = 500$\,GeV.}
\label{fig:mtbetacomb}
\end{center}
\end{figure}

Finally, we summarize our results by comparing all derived limits for the two models in the $M_{a/S_2}$--$M_A$--plane and the $M_{a/S_2}$--$\tan\beta$--plane in Fig.~\ref{fig:mMcomb} and Fig.~\ref{fig:mtbetacomb}, respectively. In addition to the $t\bar t$, mono-$Z$ and mono-$h$ constraints discussed above, we also include results from Higgs to invisible searches. The latest combined search by ATLAS~\cite{Aaboud:2019rtt} gives
\be
    \text{BR}(h \to \text{inv}) < 0.26\ \left(0.17^{+0.07}_{-0.05}\right)
\ee
for the observed (expected) upper limit at 95\,\% confidence level. As $a$ and $S_2$ dominantly decay to DM, the 3-body final state $a/S_2\, \chi\chi$ is also invisible. Using the decay widths from Sec.~\ref{sec:decays} we find a lower limit for $M_{a/S_2}$ of about $100$\,GeV for $m_\chi=10$\,GeV with a residual dependence on the mass of the heavy Higgs $M_{H/S_1} = M_A = M_{H^\pm}$. This is also the reason why we start the parameter scans at $100$\,GeV in terms of the light new Higgs mass $M_{a/S_2}$.

The differences between the pseudoscalar and scalar model in the single searches have been discussed in the corresponding last chapters, so here we focus on how the different constraints compare to each other within one model. Starting with the 2HDM+PS model in the $M_a$--$M_A$--plane with $\tan \beta = 1$, cf.~left panel of Fig.~\ref{fig:mMcomb}, we can see that the dominant limit comes from mono-$Z$ searches, excluding light Higgs masses up to $M_a \sim 320$\,GeV and heavy Higgs masses between $200$ and $1000$\,GeV, while for large masses of the heavy Higgses and small masses of the light Higgs the mono-$h$ limit leads to slightly stronger bounds. Light masses below $\sim\!100\,$GeV for the new pseudoscalar state are excluded by Higgs to invisible searches, while $t\bar t$ resonance searches can exclude small parts of the parameter space nearly independent of the light new pseudoscalar mass around  $M_{H} = M_{H^\pm} = M_A \sim 400$ and $500$\,GeV.

The same holds true for the 2HDM+S model, cf.~right panel of Fig.~\ref{fig:mMcomb}, especially for the mono-$Z$ and Higgs to invisible searches. In contrast to that, the mono-$h$ limit for the scalar model is significantly weaker compared to the mono-$Z$ limits and thus never the strongest limit, as explained in Sec.~\ref{sec:missh}. Furthermore, the $t\bar t$ limits exclude a larger band of heavy Higgs masses from $400$ to $500$\,GeV, however this is just a reflection of slightly stronger limits compared to the pseudoscalar case with the limits changing from just below $2\,\sigma$ to just above $2\,\sigma$.

For the $M_{a/S_2}$--$\tan\beta$--plane and at heavy Higgs masses of $500$\,GeV, cf.~Fig.~\ref{fig:mtbetacomb}, we have again similar limits for both models, except for the mono-$h$ limit being weaker in the scalar model. As in the $M_a$--$M_A$--plane, the dominant limit is given by mono-$Z$ searches, specifically for values of $\tan \beta > 1$. They begin to weaken for larger values of $\tan \beta$ of around three. While these limits, for $0.3\lesssim\tan\beta\lesssim 3$ apply to 2HDMs of all Yukawa sector types, cf.~Tab.~\ref{tab:coeffs}, for $\tan\beta \gtrsim3$ the $b\bar{b}$ production mode starts to become relevant for type II and Y, leading to stronger limits for such Yukawa sectors, while for type I and X $b\bar{b}$ production never becomes relevant, and limits will continue to get weaker for larger $\tan\beta$ values. The $t\bar{t}$ resonance searches provide a lower limit on $\tan \beta$ of around $1$, being slightly stronger in the 2HDM+S and nearly independent of the light mediator mass, therefore being the strongest limit for $M_{a/S_2}>270$\,GeV.

Finally, another interesting aspect which is accessible via the comparison plots is the question of how to distinguish the two models. Here, the weaker mono-$h$ limits for the 2HDM+S model can come in handy. This discrepancy in terms of sensitivity to the mono-$h$ channel could be exploited to distinguish between the 2HDM+PS and 2HDM+S, since the ratio of the signal strength in mono-$Z$ and mono-$h$ is characteristic for the model.
So if signals are detected in both channels, their signal strength ratio could be used to discriminate between the the models.

\section{Conclusions}
\label{sec:conclusions}

In this paper, we have investigated the collider phenomenology of the 2HDM+S and 2HDM+PS, in particular focusing on the $t\bar{t}$, mono-$Z$ and mono-$h$ signatures.

The 2HDM+S and 2HDM+PS are the new standard paradigm used by the CMS and ATLAS collaborations to interpret experimental results in the context of DM searches and Simplified Models. These models feature an extended scalar sector, with a second Higgs doublet, and an additional singlet and differ from the previous generation of Simplified Models and other DM models by tending to generate a rich collider phenomenology. In particular, it is possible to have resonant production of the heavy Higgs bosons that can decay to a SM boson and the additional singlet, generating resonantly-enhanced mono-$Z$ and mono-$h$ signatures.

In our analysis, we made use of a few assumptions that have been established by the experimental collaborations \citep{Abe:2018bpo} to reduce the number of dimensions of the parameter space, and simplify the comparison between the two models and the different experimental signatures. These assumptions are essentially the mass degeneracy of the heavy scalars, motivated by the bounds on the EW precision observables, together with the Higgs alignment limit, motivated by Higgs measurements. The alignment limit simplifies the Higgs sector by turning one of the doublets into the SM Higgs doublet, so that most constraints from Higgs physics are avoided. Moreover, while we have described the possible Yukawa sectors of the 2HDM, our results are universal due to the range of parameters we considered. 

We have reviewed all the principal constraints of the two models. Some of them, like Direct and Indirect Detection, are mostly complementary to collider searches, because they tend to require different mass spectra. Some other, like flavour and EW precision observables, perturbativity and unitarity constraints, instead are very relevant also for our signatures, and are used as a guidance to select the appropriate ranges for the various parameters.

Using the data provided by different LHC analyses, we derived limits for the 2HDM+S and 2HDM+PS for the $t \bar{t}$, mono-$Z$ and mono-$h$ signatures and discussed how they compare to each other and between the two models. We found that the mono-$Z$ limits are in general the most constraining ones, while the $t \bar{t}$ limits are nearly independent on the mass of the additional singlet and especially relevant for constraining $\tan \beta$. A lower bound of $\sim\!100$\,GeV for the additional singlet mass is given by Higgs to invisible searches, giving us a natural starting point for this parameter in our scans.

We found that in principle the two models could be discriminated at a collider by detecting both mono-$Z$ and mono-$h$ signatures, from the ratio of their strengths. Also, the absence or appearance of mono-jet signals would give further insights into the nature of the dark sector. However, depending on the DM mass, other probes, mostly astrophysical, would be powerful tools to help discriminating between the two models. A Direct Detection signal for DM is several orders of magnitude stronger for the 2HDM+S, while Indirect Detection signals would be several orders of magnitude stronger for the 2HDM+PS. So detecting one of these two astrophysical signatures would also give a clear indication towards the nature of the mediators to the dark sector and discriminate between 2HDM+S and 2HDM+PS. In this case, collider studies could help to further investigate the inner workings of the dark sector.

\section*{Acknowledgements}

VTT acknowledges support by the IMPRS-PTFS. The authors would like to thank Christopher Anelli, Dominick Cichon, and Stefan Vogl for valuable discussions and Tania Robens for spotting typos. Furthermore, we are grateful to Olivier Mattelaer of the MadGraph team for correspondence on the software.

\appendix

\section{Formulae for the Decay Widths}
\label{sec:Gamma_app}

In this appendix we give analytic expressions for the dominant branching ratios of the spin-$0$ states in the 2HDM+S/PS as partially shown in Sec.~\ref{sec:decays}.

We focus on the mass hierarchy $M_A,\, M_{H/S_1},\, M_{H^\pm} > M_{a/S_2},\, M_h$ and $\tan\beta =\mathcal{O}(1)$. The values of $\epsilon_f$, denoting the ratio between the Yukawa coupling of the fermion $f$ in the different types of 2HDMs and the SM value $y_f= \sqrt{2} m_f/v$, are given in Tab.~\ref{tab:coeffs}.
In the following we use the abbreviation $\tau_{i,j}=4M_i^2/M_j^2$.

\subsection{Scalar Model}
\label{sec:scalardecay}

\paragraph{Higgs Boson $h$} \label{sec:sdecayhiggs}

As mentioned above, in the decoupling limit the couplings of $h$ with the SM states substantially coincide with the ones for the SM Higgs boson. However its total width can deviate, with respect to the SM prediction, because of the eventual presence of additional decay channels. The most relevant, if kinematically allowed, is the one into a pair of $S_2$ states. As the Higgs width is small, also three-body decays to $S_2 \chi\bar\chi$ can be relevant and the additional widths are given by
\begin{align} \label{eq:GammahS}
    \Gamma(h\to S_2 S_2) &= \frac{1}{32 \pi}\ g_{hS_2 S_2}^2\, M_h \, \sqrt{1 -\tau_{S_2,h}} ,\\
    \Gamma\left(h \to S_2 \chi\bar \chi \right) &= \frac{y_\chi^2}{32 \pi^3}\ g_{hS_2 S_2}^2 \, M_h\, g(\tau_{S_2,h}) \cos^2\theta \left(1- \tau_{\chi,S_2} \right)^{3/2}, \\
    \Gamma\left(h \to S_2 f \bar f \right) &= \frac{N_c^f\, \epsilon_f^2\, y_f^2}{16 \pi^3}\ g_{hS_2 S_2}^2 \, M_h\, g(\tau_{S_2,h}) \sin^2\theta \left(1- \tau_{f,S_2} \right)^{3/2},
\end{align}
with \cite{Djouadi:1995gv}
\begin{align}\label{eq:g3}
    g(\tau) &= \frac{\tau - 4}{8} \left[4 - \ln\left(\frac{\tau}{4} \right)\right] - \frac{5\tau - 4}{4 \sqrt{\tau-1}} \left[\arctan \left(\frac{\tau-2}{2\sqrt{\tau-1}} \right) - \arctan \left( \frac{1}{\sqrt{\tau-1}} \right)\right], \\
    g_{hS_2S_2} &= \frac{1}{M_h v}\ \left(M_h^2 - 2 \,(M_{S_1}^2 -M_{S_2}^2)\cos^2\theta \right)\, \sin^2\theta .
\end{align}

\paragraph{Light Scalar $S_2$}\label{sec:sdecays2}
The light scalar $S_2$ mostly decays into $gg$, $f\bar{f}$ and $\chi\bar\chi$ (direct couplings with gauge boson are forbidden in the alignment limit), depending on the mass. We quote below the corresponding decay widths and the loop-induced one into gluons which is useful for the interpretation of the collider studies:
\begin{align} \label{eq:sGammaS2}
    \Gamma(S_2\to gg) &= \frac{\alpha_s^2}{16\pi^3}\ M_{S_2}\,\sin^2\theta\,\sum_q \epsilon_q^2\, y_q^2\,F_S\left(\tau_{q,S_2}\right),\\
    \Gamma(S_2\to f\bar{f}) &= \frac{N_c^f\, \epsilon_f^2\, y_v^2}{16\pi}\ M_{S_2}\, \sin^2\theta \left(1-\tau_{f,S_2} \right)^{3/2},\\
    \Gamma\left(S_2 \to \chi\bar\chi \right) &= \frac{y_\chi^2}{8\pi}\ M_{S_2}\, \cos^2\theta \left(1- \tau_{\chi,S_2} \right)^{3/2} ,
\end{align}
with
\begin{equation}
    F_S(x) = x \left|1+(1-x)\arctan^2\frac{1}{\sqrt{x-1}} \right|^2 .
\end{equation}

\paragraph{Heavy Scalar $S_1$}\label{sec:sdecays1}
Besides an additional $\cos^2\theta$ due to the mixing with the additional singlet the couplings of the heavy scalar to SM fields remain similar to the ones known from 2HDMs. Additional decay channels are $\chi\bar\chi$, suppressed by $\sin^2\theta$,$ S_2 S_2$, which is very small for our parameter choice and $ h S_2$ which is important for the mono-$h$ bounds. The analytic expressions are given by
\begin{align} \label{eq:GammaS1}
    \Gamma(S_1\to gg) &= \frac{\alpha_s^2}{16\pi^3}\ M_{S_1}\,\cos^2\theta\,\sum_q \epsilon_q^2\, y_q^2\,F_S\left(\tau_{q,S_1}\right),\\
    \Gamma \left(S_1 \to f \bar f \right) &= \frac{N_c^f\, \epsilon_f^2\,y_f^2}{16\pi}\ M_{S_1} \, \cos^2\theta \left(1-\tau_{f,S_1} \right)^{3/2} ,\\
    \Gamma \left(S_1 \to \chi\bar\chi \right) &= \frac{y_\chi^2}{8\pi}\ M_{S_1}\, \sin^2\theta \, \left(1- \tau_{\chi,S_1} \right)^{3/2} ,\\
    \Gamma \left(S_1 \to S_2 S_2 \right) &= \frac{1}{32 \pi}\ g_{S_1 S_2 S_2}^2\, M_{S_1}\,\sqrt{1 - \tau_{S_2,S_1}} ,\\
    \Gamma \left(S_1 \to S_2 h \right) &= \frac{1}{16 \pi}\, \frac{\lambda^{1/2}(M_{S_1}, M_h, M_{S_2})}{M_{S_1}}\ g_{S_1 h S_2}^2 ,
\end{align}
with
\begin{align}
    g_{S_1 S_2 S_2} &= \frac{1}{M_{S_1} v_S}\, \left(M_{S_1}^2 + 2 M_{S_2}^2 - \frac{2-3\sin^2\theta}{\cos^2\theta}\ \hat{\lambda}_{HHS}\, v_S^2 \right) \sin\theta \, \cos\theta ,\\
    g_{S_1 h S_2} &= \frac{1}{M_{S_1}v}\, \left(M_h^2+ \left(M_{S_1}^2- M_{S_2}^2\right) \cos2\theta\right) \sin\theta\cos\theta .
\end{align}
Furthermore, we have introduced 
\be \label{eq:lambda}
    \lambda(m_1, m_2, m_3) = \left(m_1^2 - m_2^2 - m_3^2 \right)^2 - 4\, m_2^2\, m_3^2 .
\ee

\paragraph{Pseudoscalar $A$}\label{sec:sdecayA}
Besides the partial widths known from 2HDMs the heavy pseudoscalar has an additional decay channel to $S_2Z$, which are given by
\begin{align}\label{eq:sGammaAX}
    \Gamma(A\to gg) &= \frac{\alpha_s^2}{16\pi^3}\ M_A\, \sum_q \epsilon_q^2\, y_q^2 \,F_P\left(\tau_{q,A}\right),\\
    \Gamma \left(A \to f \bar f \right) &= \frac{N_c^f\, \epsilon_f^2\, y_f^2}{16\pi}\ M_A \left(1- \tau_{f,A} \right)^{1/2} ,\\
    \Gamma \left(A \to S_2 Z \right) &= \frac{1}{16 \pi}\, \frac{\lambda^{3/2} (M_A, M_{S_2}, M_Z)}{M_A^3\, v^2},
\end{align}
with
\begin{equation} \label{eq:FP}
    F_P(x) = x \left|\arctan^2\frac{1}{\sqrt{x-1}} \right|^2 .
\end{equation}

\paragraph{Charged Scalar $H^\pm$} \label{sec:sdecayHpm}
For completeness also the partial widths of $H^\pm$ to quarks and the new spin-0 state plus a $W^\pm$, since the $H^+ h W^+$ vertex vanishes in the alignment limit, are given by
\begin{align}\label{eq:SGammaHpX}
    \Gamma \left(H^+ \to t \bar b \right ) &= \frac{N_c^t |V_{tb}|^2 \epsilon_t^2\, y_t^2}{16\pi}\ M_{H^\pm} \left(1 - \tau_{t,H^\pm}/4 \right)^2 ,\\
    \Gamma \left(H^\pm \to S_1 W^\pm \right) &= \frac{1}{16\pi}\, \frac{\lambda^{3/2} (M_{H^\pm}, M_{S_1}, M_W)}{M_{{H^\pm}}^3 v^2}\ \cos^2\theta ,\\
    \Gamma \left( H^\pm \to A W^\pm \right) &= \frac{1}{16\pi}\, \frac{\lambda^{3/2} (M_{H^\pm}, M_A, M_W)}{M_{{H^\pm}}^3 v^2} ,\\
    \Gamma \left( H^\pm \to S_2 W^\pm \right) &= \frac{1}{16\pi}\, \frac{\lambda^{3/2} (M_{H^\pm}, M_{S_2}, M_W)}{M_{{H^\pm}}^3 v^2}\ \sin^2 \theta ,
\end{align}
where in the case of $H^+ \to t \bar b$ we have neglected terms of ${\cal O}(m_b^2/M_{H^\pm}^2)$.

\subsection{Pseudoscalar Model}
\label{sec:pscalardecay}

The results in this section are taken from \cite{Bauer:2017ota} and transferred to our notation. The features are very similar to the ones in the previous section.

\paragraph{Higgs Boson $h$} \label{sec:pdecayhiggs}

Similarly to the 2HDM+S the couplings of $h$ to $\bar f f$ and gauge boson pairs are kept to their SM values by the alignment limit. Its total width might be nevertheless enlarged by additional two and/or three body decay channels being \cite{Djouadi:1995gv}:
\begin{align} \label{eq:GammahPS}
    \Gamma\left(h \to a a \right) &= \frac{1}{32 \pi}\, g_{haa}^2 \, M_h \left(1-\tau_{a,h} \right)^{1/2} , \\
    \Gamma\left(h \to a \chi\bar\chi \right) &= \frac{y_\chi^2}{32 \pi^3} \, g_{haa}^2 \, M_h\, g(\tau_{a,h}) \cos^2\theta \left(1- \tau_{\chi,a} \right)^{1/2}, \\
    \Gamma\left(h \to a f\bar f \right) &= \frac{N_c^f\, \epsilon_f^2\, y_f^2}{16 \pi^3}\ g_{haa}^2\, M_h\, g(\tau_{a,h}) \sin^2\theta \left(1- \tau_{f,a} \right)^{1/2},
\end{align}
with $g(\tau)$ given in Eq.~\eqref{eq:g3} and
\begin{align}
    g_{haa} &= \frac{1}{M_h v}\, \Big[\left(M_h^2 - 2 M_H^2 + 4 M_{H^\pm}^2 - 2 M_a^2 - 2 \lambda_3 v^2 \right) \sin^2\theta \nonumber \\
    & \hspace{1.5cm} - \left(\lambda_{11P} \cos^2\beta + \lambda_{22P} \sin^2 \beta \right) v^2 \cos^2\theta \Big] .
\end{align}

\paragraph{Light Pseudoscalar $a$} \label{sec:pdecays2}

The partial widths to $gg$ and $f\bar{f}$ and $\chi\bar\chi$ are given by
\begin{align}
    \Gamma(a\to gg) &= \frac{\alpha_s^2}{16\pi^3}\ M_a\, \sin^2\theta\, \sum_q \epsilon_q^2\, y_q^2 \,F_P\left(\tau_{q,A}\right),\\
    \Gamma\left(a\to f\bar{f}\right) &= \frac{N^f_c\, \epsilon_f^2\, y_f^2}{16\pi}\ M_a \sin^2\theta \left(1- \tau_{f,a} \right)^{1/2},\\
    \Gamma\left( a \to \chi\bar\chi\right) & = \frac{y_\chi^2}{8\pi} \ M_a \cos^2\theta \left(1- \tau_{\chi,a}\right)^{1/2},
\end{align}
where $F_P$ is given in Eq.~\eqref{eq:FP}.

\paragraph{Heavy Pseudoscalar $A$}\label{sec:pdecays1} The partial widths to $gg$ and $f\bar{f}$, $\chi\bar\chi$ and $ah$ are given by
\begin{align}\label{eq:GammaAX}
    \Gamma(A\to gg) &= \frac{\alpha_s^2}{16\pi^3}\ M_A\,\cos^2\theta\, \sum_q \epsilon_q^2\, y_q^2 \,F_P\left(\tau_{q,A}\right),\\
    \Gamma \left( A \to f \bar f \right) &= \frac{N_c^f\, \epsilon_f^2\, y_f^2}{16\pi}\ M_A\, \cos^2 \theta \left(1- \tau_{f,A} \right)^{1/2} ,\\
    \Gamma \left( A \to \chi \bar \chi \right) & = \frac{y_\chi^2}{8\pi}\ M_A\, \sin^2\theta \left(1- \tau_{\chi,A} \right)^{1/2},\\
    \Gamma \left(A \to a h \right) &= \frac{1}{16 \pi}\, \frac{\lambda^{1/2} (M_A, M_a, M_h)}{M_A}\ g_{Aah}^2 ,
\end{align}
with
\begin{align}
\label{eq:gAah}
    g_{Aah} = \frac{1}{M_A\, v} \, &\Big[M_h^2 - 2 M_H^2 - M_A^2 + 4 M_{H^\pm}^2 - M_a^2 \\
    &+\, \left( \lambda_{11P} \cos^2 \beta + \lambda_{22P} \sin^2 \beta - 2 \lambda_3 \right) v^2 \Big ] \sin\theta \cos\theta . \nn
\end{align}

\paragraph{Heavy Scalar $H$}\label{sec:pdecayA} The partial widths to $gg$ and $f\bar{f}$, $aa$ and $aZ$ are given by
\begin{align} \label{eq:GammaH}
    \Gamma(H\to gg) &= \frac{\alpha_s^2}{16\pi^3}\ M_H\, \sum_q \epsilon_q^2\, y_q^2\,F_S\left(\tau_{q,H}\right) ,\\
    \Gamma \left(H \to f\bar f\right) &= \frac{N_c^f\, \epsilon_f^2\,y_f^2}{16\pi}\ M_H \left(1-\tau_{f,H} \right)^{3/2} ,\\
    \Gamma \left(H \to a a \right) &= \frac{1}{32 \pi}\ g_{Haa}^2\, M_H \left(1-\tau_{a,H} \right)^{1/2} ,\\
    \Gamma \left(H \to a Z \right) &= \frac{1}{16 \pi}\, \frac{\lambda^{3/2} (M_H, M_a, M_Z)}{M_H^3 v^2}\ \sin^2 \theta ,
\end{align}
with
\begin{align}\label{eq:gHaa}
    g_{Haa} = \frac{1}{M_H v}\, &\Big[\cot \left( 2 \beta \right) \left(2 M_h^2 - 4 M_H^2 + 4 M_{H^\pm}^2 - 2 \lambda_3 v^2 \right) \sin^2 \theta\\\nn
    &+ \sin\left(2 \beta \right) \cos^2\theta\,v^2\, \left(\lambda_{11P}-\lambda_{22P} \right)/2\Big] ,
\end{align}
denoting the $Haa$ coupling and $\lambda$ is given in Eq.~\eqref{eq:lambda}.

\paragraph{Charged Scalar $H^\pm$}\label{sec:pdecayHpm}

Since in the alignment limit the $H^+ h W^+$ vertex vanishes, the partial decay widths of the charged scalar $H^\pm$ relevant for small $\tan\beta$ are given by
\begin{align}\label{eq:GammaHpX}
    \Gamma \left(H^+ \to t \bar b \right ) &= \frac{N_c^t\, |V_{tb}|^2\, \epsilon_t^2\, y_t^2}{16\pi}\ M_{H^\pm} \left(1 - \tau_{t,H^\pm}/4 \right)^2 ,\\
    \Gamma \left(H^\pm \to H W^\pm \right) &= \frac{1}{16\pi}\, \frac{\lambda^{3/2} (M_{H^\pm}, M_H, M_W)}{M_{{H^\pm}}^3 v^2} ,\\
    \Gamma \left(H^\pm \to A W^\pm \right) &= \frac{1}{16\pi}\, \frac{\lambda^{3/2} (M_{H^\pm}, M_A, M_W)}{M_{{H^\pm}}^3 v^2}\ \cos^2\theta ,\\
    \Gamma \left(H^\pm \to a W^\pm \right) &= \frac{1}{16\pi}\, \frac{\lambda^{3/2} (M_{H^\pm}, M_a, M_W)}{M_{{H^\pm}}^3 v^2}\ \sin^2 \theta ,
\end{align}
where in the case of $H^+ \to t \bar b$ we have neglected terms of ${\cal O}(m_b^2/M_{H^\pm}^2)$ again.

%%%%%%%%%%%%%%%%%%%%%%%%%%%%%%%%%%%%%%%%%%
%	BIBLIOGRAPHY
%%%%%%%%%%%%%%%%%%%%%%%%%%%%%%%%%%%%%%%%%%

\label{Bibliography}
\lhead{\emph{Bibliography}} % Change the page header to say "Bibliography"

\bibliography{Bibliography}

\end{document}